\documentclass[12pt]{iopart}
\usepackage[utf8]{inputenc}
\usepackage{geometry}
\usepackage{amsbsy}
\usepackage{graphicx}
\usepackage{setspace}
\usepackage{esint}
\renewcommand{\vec}[1]{{\bf #1}}
\newcommand{\eqref}[1]{(\ref{#1})}
\makeatletter
\usepackage[pdftex]{hyperref}
\usepackage{cite}
\usepackage{etoolbox}

\newcommand{\grad}\nabla
\makeatother
\usepackage{wasysym}
\usepackage{tikz}
\newcommand{\bff}{ {\bf f}}
\newcommand{\bfr}{ {\bf r}}
\newcommand{\bfv}{ {\bf v}}
\newcommand{\bfu}{ {\bf u}}
\newcommand{\bfp}{ {\bf p}}
\newcommand{\bfdp}{ {\bf \Delta p}_i^{\rm a}}
\newcommand{\bfdpf}{ {\bf \Delta p}^{\rm a}}
\newcommand{\eps}{\varepsilon}

\makeatletter
\def\@mkboth#1#2{}
\newlength\appendixwidth
\preto\appendix{\addtocontents{toc}{\protect\patchl@section}}
\newcommand{\patchl@section}{%
  \settowidth{\appendixwidth}{\textbf{Appendix }}%
  \addtolength{\appendixwidth}{1.5em}%
  \patchcmd{\l@section}{1.5em}{\appendixwidth}{}{\ddt}%
  \patchcmd{\l@subsection}{2.3em}{\appendixwidth}{}{\ddt}%
}
\makeatother

\begin{document}

\date{today}
\title{Mechanical pressure and momentum conservation in dry active matter}

\author{Y. Fily}
\address{Martin Fisher School of Physics, Brandeis University, Waltham, MA 02453, USA}
\author{Y. Kafri}
\address{Department of Physics, Technion, Haifa 32000, Israel}
\author{A. Solon}
\address{Department of Physics, Massachusetts Institute of Technology, Cambridge, Massachusetts 02139, USA}
\author{J. Tailleur\footnote{Invited author}}
\address{Universit\'e Paris Diderot, Sorbonne Paris Cit\'e, MSC, UMR 7057 CNRS, 75205 Paris, France}
\author{A. Turner}
\address{Department of Physics, Technion, Haifa 32000, Israel}

\begin{abstract}
  We relate the breakdown of equations of states (EOS) for the
  mechanical pressure of generic dry active systems to the lack of
  momentum conservation in such systems. We show how net sources and sinks
  of momentum arise generically close to confining walls. These
  typically depend on the interactions of the container with the
  particles, which makes the mechanical pressure a container-dependent
  quantity. We show that an EOS is recovered if the dynamics of the
  propulsive forces of the particles are decoupled from other degrees of
  freedom and lead to an apolar bulk steady-state. This recovery of an
  EOS stems from the mean steady-state active force density being the
  divergence of the flux of ``active impulse'', an observable which
  measures the mean momentum particles will receive from the substrate
  in the future.
\end{abstract}
\maketitle

%

\tableofcontents{}

~\newpage

Active particles convert energy stored in the environment into
self-propelling mechanical forces. They have attracted a lot of
interest in recent years~\cite{Marchetti2013RMP} because of their
broad applicability to model systems ranging from
biological~\cite{Julicher2007PhysRep} through
granular~\cite{Narayan:2007:Science,Deseigne2010PRL}, to colloidal
systems~\cite{Palacci:2010:PRL,Galajda2007,Bricard2013Nature}. They
also hold tremendous promise for making great micro-scale
machines~\cite{DiLeonardo2010PNAS}.  A key aspect of designing such
machines is predicting the mechanical pressure an active system exerts
on its boundaries, a topic that is attracting much theoretical
interest~\cite{Yang2014,Mallory2014PRE,Takatori2014,Fily2014a,Solon:2015:NatPhys,Solon_interactions,Ginot:2015:PRX,Takatori2015,Yan2015JFM,Yan2015SoftMatter,Winkler2015SoftMatter,Speck:2016:PRE,Nikolai2016PRL,Joyeux:2016:PRE,Falasco:2016:NJP,Steffenoni:2016:arxiv,Marchetti2016,Junot2017}.

In standard thermodynamics, the pressure obeys an equation of state
(EOS): it only depends on the bulk properties of the system and is
independent of its boundary conditions. This has many
implications. Consider for example a cavity filled with a fluid and
separated into two halves by a mobile piston. The piston can only be
set in motion if the fluids on each side have different bulk
properties (e.g. densities or temperatures).  This in turn constrains
engine designs.

In contrast, we recently showed~\cite{Solon:2015:NatPhys} that the
pressure exerted by an active fluid on a flat wall need not obey an
EOS. It generally depends explicitly on the potential describing the
wall. Coming back to our container separated in two parts by a mobile
piston, motion can now arise even when the fluids on both sides of the
piston have the same bulk properties, provided the piston's surfaces
are different~\cite{Solon:2015:NatPhys}. This wall dependence has
since been shown experimentally in shaken granular
systems~\cite{Junot2017}. Importantly, Refs.~\cite{Solon:2015:NatPhys}
and~\cite{Junot2017} deal with so-called \emph{dry active systems},
i.e., systems that do not obey detailed balance (as is generically the
case for active systems) and have no local momentum conservation (by
pushing on a substrate or surrounding medium which acts as a momentum
sink).

Note, however, that some dry active systems have been
shown to admit an
EOS~\cite{Takatori2014,Yang2014,Solon:2015:NatPhys,Solon_interactions}.
The role of pressure in such systems then shares similarities with its
role in standard thermodynamics: for instance, it is equal in
coexisting phases\cite{Solon_interactions}. It can also be used to
define an isobaric ensemble and hence to control active
systems~\cite{Solon:arxiv:2016}. However, this link to equilibrium
physics is only partial and, for instance, the Maxwell construction in
the pressure-volume phase diagram does not yield the correct binodals
in phase-separating active systems~\cite{Solon_interactions}.

The goal of this paper is to relate the lack of EOS to a violation of
momentum conservation. To do so, we revisit the results of
Ref.~\cite{Solon:2015:NatPhys} for a class of microscopic models of
active particles that explicitly account for their translational
inertia, instead of studying the commonly used overdamped models. (Our
previous results are then naturally recovered in the large damping
limit.).

To help contextualize our results and state them more accurately, it
is useful to first consider an equilibrium system and then a system
(at equilibrium or not) with local momentum conservation.  In the
former, the extensivity of the free energy alone ensures the pressure
only depends on bulk properties.  In the latter, the momentum density
field ${\bf p}$ obeys the conservation equation~\cite{irving1950}
\newcommand{\ha}{}
\begin{equation}\label{eq:boring}
\partial_t {\ha{\bf p}} = -\nabla \cdot {\bf J}_{\ha{\bf p}}\;.
\end{equation}
Here ${\bf J}_{\ha {\bf p}}$ is a tensorial current associated with
the local conservation law for the momentum density (the flow of
momentum for non-interacting particles). In the presence of an external wall, interacting with the particles through a potential $V_{\rm ext}$, momentum can be exchanged with the system and the dynamics~\eqref{eq:boring} becomes
\begin{equation}\label{eq:conservation}
\partial_t {\ha{\bf p}} = -\nabla \cdot {\bf J}_{\ha{\bf p}} - \ha\rho
\nabla V_{\rm ext}\;,
\end{equation}
where $\ha\rho$ is the particle density. The last term describes the exchange of momentum with the wall. 
Assuming for simplicity that the wall is flat and oriented normal to
the $x$--direction (see Fig.~\ref{fig:walls}), the pressure exerted on
the wall is simply given by 
\begin{equation}
P = \int_{x_b}^\infty dx  \ha\rho\, \partial_x V_{\rm ext}(x) 
\label{eq:pressure_definition}
\end{equation}
where $x_b$ refers to a point in the bulk, far away from the
wall. Eqs.~\eqref{eq:conservation}-\eqref{eq:pressure_definition} are
exact provided $\ha\rho$, ${\ha{\bf p}}$, and ${\bf J}_{\ha{\bf p}}$
are understood as local statistical averages of the underlying
instantaneous fields. At steady-state,
substituting~\eqref{eq:conservation}
into~\eqref{eq:pressure_definition} readily yields $P=\left({\bf
  J}^{xx}_{\ha {\bf p}}\right)_{x=x_b}$.  Assuming that the walls have
no influence on the properties of the fluid in the bulk of the system,
$\left({\bf J}^{xx}_{\ha{\bf p}}\right)_{x=x_b}$ is independent
of the choice of $x_b$ and $P$ is a bulk
property, independent of the specific wall potential $V_{\rm ext}$.

In dry active systems, the self-propulsion force acts as a local
source of momentum field on the right-hand-side of
Eq.~\eqref{eq:conservation}. Since its spatial integral
has no reason to be wall independent, there is no reason for there to
be an equation of state.  In this paper, we show how to break down the
active force field into equation-of-state-breaking and
equation-of-state-nonbreaking terms, what aspect of the particle
dynamics gives rise to each, and how to interpret them.
In section~\ref{sec:NItorques} we first consider the case of
non-interacting particles. We show how torques exerted by the walls on
the particles generically result in net steady-state sources or sinks
of momentum localized close to the confining walls
(sections~\ref{sec:walltorques} and~\ref{sec:numericstorques}). These
momentum-non-conserving terms are wall-dependent and lead to a lack of
an equation of state. Conversely, in the absence of such terms in the
steady state, we introduce an effective momentum which includes a
novel \emph{active impulse} term, defined as the mean momentum a
particle will receive on average from the substrate in the future
(section~\ref{sec:impulse}), and related to the well-known swim
pressure (section~\ref{app:swimpressure}). This effective momentum
satisfies a conservation law in the steady state, which restores the
equation of state for the pressure of
Refs.~\cite{Takatori2014,Yang2014} (section~\ref{app:momcon}).
In section~\ref{sec:interactions} we
generalize this discussion to the case of interacting particles and
show how the same reasoning allows one to understand the emergence of
an equation of state for pairwise forces and its absence for quorum-sensing interactions. We then discuss these results in the context of
Motility-Induced Phase Separation (MIPS)~\cite{Cates2015MIPS} in
section~\ref{sec:MIPS}.

\section{Noninteracting active particles}
\label{sec:NItorques}

\begin{figure}
  \begin{center}
  \begin{tikzpicture}
    \shadedraw[white,left color=gray,right color=white] (0,0) rectangle (1,3);
    \shadedraw[white,left color=white,right color=gray] (5,0) rectangle (6,3);
    \draw[ultra thick, black] (0,0) -- (0,3);
    \draw[ultra thick, black] (6,0) -- (6,3);
    \draw[ultra thick, black, dashed] (0,0) -- (6,0);
    \draw[ultra thick, black,dashed] (0,3) -- (6,3);

    \foreach \i in {1,2,...,30}{
      \begin{scope}[xshift=rand*2.75cm+3cm,yshift=rand*1.25cm+1.5cm]
        \fill[red] (0,0) circle (.05);
        \draw[red,thick,rotate=rand*180+180,scale=.3,->] (0,0) -- (1,0);
      \end{scope}}

    \draw[ultra thick, gray, dashed] (3,-.25) node[below] {$x_b$} -- (3,3.25);

    \hspace{.5cm}
    
    \begin{scope}[xshift=8cm,scale=.6]
      \draw[->,line width=0.05cm] (-1,0) --(7.5,0) node[below] {$x$};
      \draw[->,line width=0.05cm] (0,-1) --(0,4);
      \draw[red,line width=0.09cm,dashed] (0,0) -- (2.2,0); 
      \draw[red,line width=0.07cm] (2.2,0) -- (3.2,0); 
      \draw[red,line width=0.07cm,domain=3.2:6.4] plot
(\x,{0.4*(\x-3.2)*(\x-3.2)}); 
      \draw[red] (7,3.5) node { $V_{\rm ext}$};
      \draw[blue,line width=0.07cm] (0,2) -- (2,2); 
      \draw[blue,line width=0.07cm] (2,2) -- (3,2); 
      \draw[blue,line width=0.07cm,dashed,domain=3:7] plot
(\x,{2*exp(-0.4*(\x-3)*(\x-3))}); 
      \draw[blue,line width=0.07cm,dashed,domain=3:7] plot
(\x,{2*(1+(\x-3)*(\x-3))*exp(-0.4*(\x-3)*(\x-3))}); 
      \draw[blue] (1.95,3) node {$\rho(x,0)$};
      \draw [ultra thick, gray] (.5,-.25) node[below] {$x_b$}-- (.5,.25);
    \end{scope}
  \end{tikzpicture}
  \caption{We consider a simple setting with vertical, flat, walls
    confining the particles along the $\hat x$ direction while
    periodic boundary conditions are used along the $\hat y$
    direction. Active particles are unaffected by the interactions
    with the walls in the bulk of the system (white background). In
    the gray region, active particles experience a repulsive potential
    from the wall, which makes their density vanish at the wall
    boundary. On the right, two sketches of the density profiles are
    shown, with or without accumulation of active particles depending
    on the details of the potential and of the active dynamics. $x_b$
    refers to any position deep in the bulk of the system, far away
    from the walls.}
\end{center}
\label{fig:walls}
\end{figure}
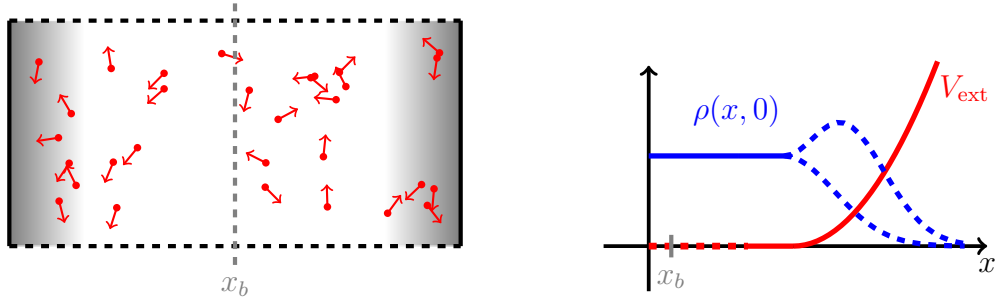

We first consider non-interacting active particles confined between
two flat walls (see Fig.~\ref{fig:walls}). For the sake of
concreteness, we first derive our results in the case of an
underdamped active Brownian particle model, described in
section~\ref{zemodel}, before generalizing to a much wider class of
models in~\ref{sec:SCEOS}.

\subsection{The Model}
\label{zemodel}
We consider active particles evolving in two spatial
dimensions. Particle $i$, located at ${\bf r}_i$, evolves according to
the dynamics:
\begin{eqnarray}
\dot{\vec r}_i&=\vec v_i\nonumber \\
m\dot{\vec v}_i&=-\tilde{\gamma} \vec v_i +f_i \vec u (\theta_i)-\vec\nabla_{{\bf r}_i}
V_{\rm ext}(\bfr_i) +\sqrt{2 \tilde{\gamma}^2 D_t} \boldsymbol{\eta}_i
\;.\label{eq:dyn}
\end{eqnarray}
Here $\tilde\gamma$ is the friction coefficient, or inverse mobility,
of the active particles, $f_i$ their propulsive forces, $V_{\rm ext}$
is the potential exerted by the confining walls, and
$\boldsymbol{\eta}_i$'s are Gaussian white noises satisfying $\langle
{\eta}^\alpha_i(t) \rangle=0$ and $\langle {\eta}^\alpha_i(t)
{\eta}^\beta_j(t') \rangle = \delta_{i,j}\delta_{\alpha,\beta}
\delta(t-t')$.  Angular brackets denote an average over noise
realizations. ${\bf u}(\theta_i)$ is a unit vector along the
orientation $\theta_i$ of particle $i$. It evolves according to the
overdamped dynamics
\begin{equation}\label{eq:theta}
  \dot{\theta}_i= \Gamma_i({\bf r}_i,\theta_i)  + \sqrt{2D_r} \zeta_i
\end{equation}
where $\Gamma_i({\bf r}_i,\theta_i)$ is the torque exerted on the
particle by the wall scaled by the rotational mobility, and $\zeta_i$
is a Gaussian white noise with $\langle \zeta_i(t) \rangle=0$ and
$\langle \zeta_i(t)\zeta_j(t')\rangle = \delta_{i,j}\delta(t-t')$.
Note that we consider underdamped dynamics, i.e. we retain the
translational inertia of the particles, instead of the standard
overdamped dynamics that we studied previously~\cite{Solon:2015:NatPhys}
\begin{equation}
   \vec v_i = \frac{1}{\tilde{\gamma}} f_i \vec u (\theta_i)-\frac{1}{\tilde{\gamma}} \vec\nabla_{{\bf r}_i}
  V_{\rm ext}(\bfr_i) +\sqrt{2 D_t} \boldsymbol{\eta}_i \;.
\end{equation}

A system of $N$ such particles can always be characterized by its
exact microscopic number density and momentum density fields
\begin{equation}\label{eq:momentum}
\hat
\rho(\vec r)=\sum_{i=1}^N \delta(\vec r-\vec r_i)\qquad\hat{\vec p}(\vec r)=\sum_{i=1}^N  m \vec v_i \delta (\vec r-\vec r_i)\;,
\end{equation}
These fields are fluctuating quantities whose averages with respect to
noise realizations and initial conditions are defined by
\begin{equation}
  \vec p(\vec r)=\langle \hat {\vec p}(\vec r) \rangle \qquad \rho(\vec
  r)=\langle \hat \rho(\vec r)\rangle.
\end{equation}

\subsection{Momentum and Pressure}\label{sec:mompress}

To proceed, we consider the dynamics of the momentum and density
fields. The density field evolves according to
\begin{equation}\label{eq:dynrho}
 \partial_t {\hat \rho}(\vec r) = \sum_i  \dot {\vec r}_i \cdot \grad_{\vec
r_i}\delta(\vec r-\vec r_i) = -\frac{1}{m}\grad \cdot \hat {\vec p}(\vec r)
\;.
\end{equation}
Here the subscript $\vec r_i$ signals that the first gradient
acts on the coordinates of particles $i$ whereas the absence of a subscript in the last
divergence signals that it acts on the position $\vec r$ where the density is
measured. This notation is used throughout the paper. Similarly, the
dynamics of the momentum density field read
\begin{eqnarray}
\label{mom_dynamics}
\partial_t \vec{\hat p} &= \sum_i\left( -\tilde{\gamma} {\vec v}_i
  -\nabla_{\vec r_i} V_{\rm ext} +f_i \vec u(\theta_i) +\sqrt{2 \tilde{\gamma}^2 D_t}
  \eta_i\right)\delta(\vec r-\vec r_i) +\sum_i  m \vec {v_i} (\vec{v_i} \cdot
\grad_{\vec r_i}) \delta(\vec r-\vec r_i)  \nonumber \\
&= -\gamma \vec{\hat p} -  \hat \rho
\nabla V_{\rm ext}+\sum_i  f_i \vec
u(\theta_i) \delta(\vec r-\vec r_i) +\sqrt{2 \tilde{\gamma}^2 D_t \hat \rho} \,\boldsymbol{\Lambda}      -
\grad\cdot [  {\cal J} ] \;.
\end{eqnarray}
Here $\gamma\equiv \tilde{\gamma}/m$, and the tensor ${\cal J} $ is
defined by
\begin{equation}\label{eq:defJ}
  {\cal J} \equiv
  \sum_i m{\vec v_i}{\vec v_i} \delta(\vec r-\vec r_i)
\end{equation}
where ${\bf v}_i {\bf v}_i$ implies a tensor product\footnote{Note
  that $\langle {\cal J} \rangle$ is non-zero even in steady-state due
  to the trivial correlations between velocities and momenta.}. The
$(i,j)$ component of the tensor ${\cal J}$ is the flux along $\hat
\imath$ of momentum along $\hat \jmath$. In addition, we have defined
the Gaussian white noise
$\displaystyle\sqrt{2 \tilde{\gamma}^2 D_t \hat \rho} \,\boldsymbol\Lambda\equiv \sum_i \sqrt{2 \tilde{\gamma}^2 D_t} \boldsymbol{\eta}_i
\delta({\bf r}-{\bf r_i})$
which can be verified to obey $\langle \Lambda_\alpha({\bf r},t)
\Lambda_\beta({\bf r'},t')\rangle = \delta_{\alpha \beta}\delta({\bf
  r}-{\bf r'})\delta (t-t')$. Eq.~\eqref{mom_dynamics} lists the
various contributions leading to momentum density changes in ${\bf r}$: (i) loss of momentum due to dissipation, $-\gamma \vec{\hat
  p}$; (ii) forces due to the walls, $- \hat \rho \nabla V_{\rm
  ext}$; (iii) active forces propelling the particles, $\sum_i f_i
\vec u(\theta_i) \delta(\vec r-\vec r_i)$; (iv) fluctuations, $\sqrt{2
  \tilde{\gamma}^2 D_t \hat \rho}\, {\bf \Lambda}$; (v) advection of momentum as
particles arrive in and depart from $\vec r$, $ - \grad\cdot [ {\cal
  J} ] $.

At steady state, in the confining potential shown in
Fig.~\ref{fig:walls}, Eq.~\eqref{eq:dynrho} implies that ${\bf p}(\vec
r)$=0. Since the system is invariant under translations along $\hat y$
we integrate Eq.~\eqref{mom_dynamics} along the $y$ coordinate and
define $\rho(x)=\int_0^1 dy \rho (x,y)$. Here, to keep the notation
simple, we set the extent of the system in the $y$ direction to be
unity and we silently omit the (lack of) dependence on $y$ of the
observables. This gives
\begin{equation}\label{eq:localpressure}
  0 = -\rho(x) \partial_x V_{\rm ext}(x)+    \langle\sum_i  f_i \cos \theta_i
  \delta(x-x_i)  \rangle - \partial_x[ \langle {\cal J}^{xx}(x) \rangle ]
\end{equation}
where ${\cal J}^{xx}=\sum_i \delta(x-x_i) m \vec
(v^x_i)^2$. Eq.~\eqref{eq:localpressure} states that the momentum
flux is enhanced or suppressed as it travels through the system by
either the wall force or the active forces.

The mechanical pressure exerted on the wall is then given by integrating 
Eq.~\eqref{eq:localpressure} from a point $x_b$ deep in the bulk of the system to $x=+\infty$
\begin{equation}\label{eq:globalpressure}
 P\equiv \int_{x_b}^\infty dx \rho \nabla_x V_{\rm ext} =  \langle {\cal J}^{xx}({x_b}) \rangle 
+  \int_{x_b}^{\infty} \langle\sum_i  f_i \cos \theta_i \delta(x-x_i) \rangle dx
\;.
\end{equation}
The total change of the momentum flux from its bulk value $\langle
{\cal J}^{xx}\rangle$ to zero beyond the wall is the result of the
total force exerted by the wall and of the total active force exerted
in the $x>{x_b}$ region. Alternatively, Eq.~\eqref{eq:globalpressure}
means that the pressure is equal to the momentum flux entering the
$x>{x_b}$ region plus the total active force exerted in this
region. Note that, since $\grad_x V_{\rm ext}$ vanishes in the bulk,
$P$, as defined in~\eqref{eq:globalpressure}, does not depend on the
choice of $x_b$. 

To have an EOS, the right-hand side of
equation~\eqref{eq:globalpressure} has to solely depend on bulk
quantities. This is clearly the case for $\langle {\cal J}^{xx}({x_b})
\rangle$ which originates from the divergence of the momentum flux.
On the contrary, the total active force exerted in the $x>{x_b}$
region has no reason to depend only on bulk properties.  While in the
bulk the isotropy of the steady-state implies that $ \langle\sum_i f_i
\cos \theta_i \delta(x-x_i) \rangle$ is zero, it is well known that
near the wall a non-trivial orientational order develops which depends
explicitly on the potential shape $V_{\rm ext}$. It is thus natural to
expect that the mean active force should generically give a
wall-dependent contribution to the pressure, hence leading to a lack
of equation of state. Note that neither the damping force $-\gamma
\hat p$ nor the Langevin force $\sqrt{2 \tilde{\gamma}^2 D_t \hat
  \rho} {\bf\Lambda}$ conserve momentum either. However, they average to zero
in steady state and therefore do not affect the pressure directly.

Expectedly, setting $f_i=0$ in our model gives back the equilibrium
dynamics of underdamped colloidal
particles. Eq.~\eqref{eq:globalpressure} then expresses the mechanical
pressure in terms of a purely bulk quantity, $\langle {\cal
  J}^{xx}({x_b}) \rangle $; this fleshes out the argument presented in
the introduction after Eq.~\eqref{eq:pressure_definition} and shows
such equilibrium systems to admit an EOS.

We next illustrate how the active force influences the pressure for
the system described by Eqs.~\eqref{eq:dyn} and~\eqref{eq:theta}.

\subsection{The  breakdown of the EOS: momentum sources and sinks}\label{sec:walltorques}
Using It\=o calculus, the dynamics of the mean active force density
is given by
\begin{eqnarray}\label{eq:localrateactiveforce}
  \partial_t \langle \sum_i f_i \cos\theta_i \delta(x-x_i) \rangle &=  -
\langle \sum_i f_i \Gamma_i \sin\theta_i \delta(x-x_i) \rangle - D_r \langle
\sum_i f_i \cos\theta_i \delta(x-x_i) \rangle\nonumber\\
  &\qquad  -
\partial_x[ \langle \sum_i v_i^x f_i \cos\theta_i \delta(x-x_i) \rangle]
\end{eqnarray}
which in the steady state, gives the balance equation
\begin{equation}\label{eq:localmeanactiveforce}
  \langle \sum_i f_i \cos\theta_i \delta(x-x_i) \rangle =  -
\langle \sum_i \frac{f_i}{D_r} \Gamma_i \sin\theta_i \delta(x-x_i)
\rangle   - \partial_x[ \langle \sum_i \frac{v_i^x}{D_r} f_i
\cos\theta_i \delta(x-x_i) \rangle]\;.
\end{equation}
This states that the local active force is the sum of a
torque-dependent term and a torque-independent one, the latter being
the derivative of a \textit{local} quantity. This local quantity is
responsible for the advection of active forces and we discuss in
details its physics in section~\ref{sec:impulse}.

The force balance equation~\eqref{eq:localpressure} now becomes
\begin{equation}\label{eq:forcebalance}
  \rho(x) \partial_x V_{\rm ext}(x)  = - \sum_i  \langle\frac{f_i}{D_r} \Gamma_i \sin\theta_i \delta(x-x_i)
  \rangle 
  - \partial_x\Big [ \sum_i \langle \frac{v_i^x}{D_r} f_i
  \cos\theta_i \delta(x-x_i) \rangle+ \langle {\cal J}^{xx}(x) \rangle \Big] \;.
\end{equation}
{Eq.~\eqref{mom_dynamics} identified the active forces as
  momentum sources, in agreement with the idea that each particle is
  receiving momentum from the
  substrate. Eqs.~\eqref{eq:localmeanactiveforce}
  and~\eqref{eq:forcebalance} show that the steady-state mean active
  force density can be split between a ``momentum-conserving'' part,
  \textit{i.e.} the divergence of a local tensor, and a
  torque-dependent ``non-conserving'' term, i.e. a steady-state momentum
  source or sink. Note that this ``effective momentum conservation''
  is restricted to the steady-state and we discuss it in more detail in
  section~\ref{sec:impulse} and~\ref{app:momcon}. For now, we focus
on the pressure, which can be written as}
\begin{equation}\label{eq:pressureEOStorque}
 P = \langle {\cal J}^{xx}({x_b}) \rangle  + \sum_i \langle \frac{v_i^x}{D_r}
f_i \cos\theta_i \delta({x_b}-x_i)\rangle
-\int_{x_b}^\infty dx \langle \frac{f_i}{D_r} \Gamma_i \sin \theta_i \delta (x-x_i)
\rangle \;,
\end{equation}
where ${x_b}$ is a point deep in the bulk of the system. The first two
terms depend only on bulk properties of the fluid, while the latter is
responsible for a wall-dependent contribution to the pressure. The
breakdown of the equation of state thus arises from wall-dependent
active sources. Eq.~\eqref{eq:localmeanactiveforce} shows that these
sources can be measured by removing the contribution of the advective
active-force term from the total contribution of the active
force. Namely, by looking at
\begin{equation}\label{eq:source}
\Delta f_{\rm act}(x)\equiv  \langle \sum_i f_i \cos \theta_i \delta(x-x_i) \rangle +\partial_x \langle \frac{v_i^x}{D_r} f_i \cos\theta_i \delta(x-x_i)\rangle
  \rangle\,.
\end{equation}
The relation is, as will become clearer later, more general and can be
used to identify the presence of steady-state momentum sources for any
model of active Brownian particles. For the model we consider here, in
the presence of wall torques, the non-conserving terms are given by
\begin{equation}
\Delta f_{\rm act}(x)=-\sum_i  \langle\frac{f_i}{D_r} \Gamma_i \sin\theta_i \delta(x-x_i)\rangle\;.
\end{equation}

\begin{figure}
  \begin{center}
    \includegraphics[width=.9\textwidth]{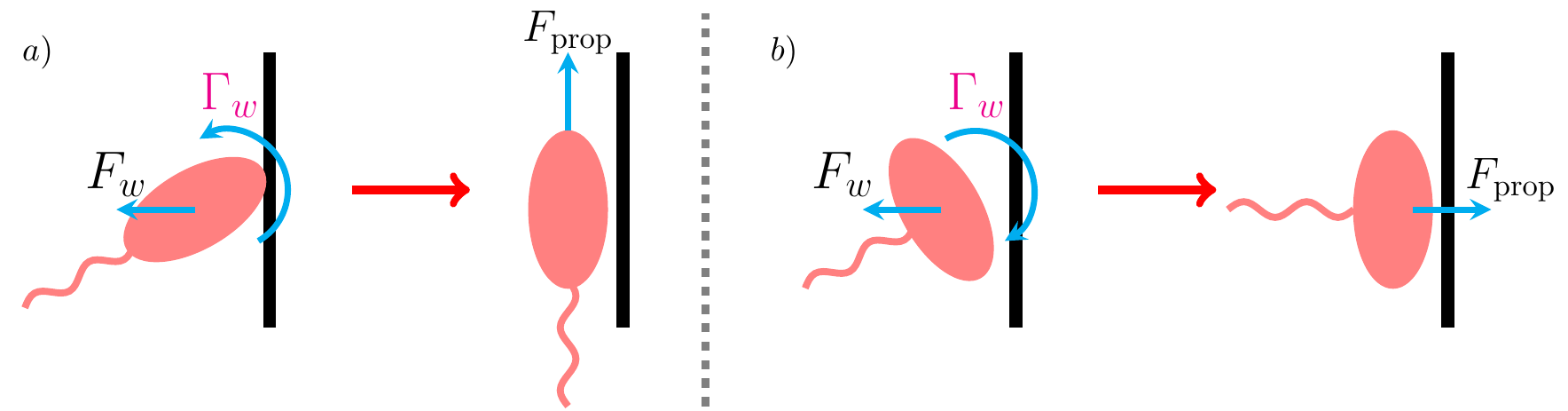}
  \end{center}
  \caption{{Impact of wall torques on the active contribution to the
      mechanical pressure}. {\bf $a)$} Torques that make active particles align
    their propulsive forces along the wall diminish their
    contributions to the overall force exerted on the wall. {\bf $b)$} On
    the contrary, torques making the active particles face the wall
    increase the contributions of the active forces to the mechanical
    pressure.}
  \label{fig:figtorque}
\end{figure}

Figure~\ref{fig:figtorque} offers a more intuitive picture of the way
wall torques affect the pressure. The active pressure comes from the
transmission of the momentum received by the particles from the
substrate to the walls. If torques are such that particles align and
move along the wall, the momentum they transmit to the wall is reduced
(compared to say, torque-less particles). Conversely, if torques force
the particle to face the walls, the transmission of active force to
the wall is enhanced.

All in all, Eq.~\eqref{eq:globalpressure} shows the pressure to be
related to the mean active forces experienced by the particles. Even
though each of these forces injects momentum into the system,
Eq.~\eqref{eq:localmeanactiveforce} shows the mean active force in
steady state to be composed of two contributions: a torque-dependent
source, measured in Eq.~\eqref{eq:source}, and a momentum-conserving
one. Before focusing on the torque-free case
(section~\ref{sec:impulse}) and showing how to interpret those results
in terms of an effective momentum conservation
(section~\ref{app:momcon}), let us numerically illustrate our results
so far.

\subsection{Numerical measurement of sources and sinks}
\label{sec:numericstorques}

We consider self-propelled particles evolving under the
dynamics~\eqref{eq:dyn} and~\eqref{eq:theta}, confined by harmonic
walls modelled by the repulsive potentials
\begin{equation}
\label{eq:potential}
  V^R(x)=\lambda_R\frac{(x - x_w)^2}2\Theta(x-x_w)\qquad\mbox{and}\qquad   V^L(x)=\lambda_L\frac{(x + x_w)^2}2 \Theta(-x-x_w)
\end{equation}
for right and left walls, respectively. Here $\Theta(x)$ denotes a
Heaviside function. The particles are modeled as ellipses of principle
axes $a$ and $b$, with the self-propulsion being along the $a$
axis. We consider the limit in which the particles are `point-like'
(\textit{i.e.} much smaller than their penetration length into the wall
potential) so that the torques they experience from the walls can be
computed explicitly as
\begin{equation}
  \Gamma^R(x,\theta)=\lambda_R\,\kappa \Theta(x-x_w) \sin 2\theta\qquad\mbox{and}\qquad  \Gamma^L(x,\theta)=\lambda_L\,\kappa \Theta(-x-x_w) \sin 2\theta
\end{equation}
where $\kappa=\mu_r(a^2-b^2)$ measures the anisotropy of the particles
and $\mu_r$ is a rotational mobility~\cite{Solon:2015:NatPhys}. In
practice, we took $x_b=0$ in all our simulations (See
Eq.~\ref{eq:globalpressure}), making sure that the system size was
large enough that a moderate change in $x_b$ did not change our
results.

\begin{figure}[h]
\includegraphics[width=\textwidth]{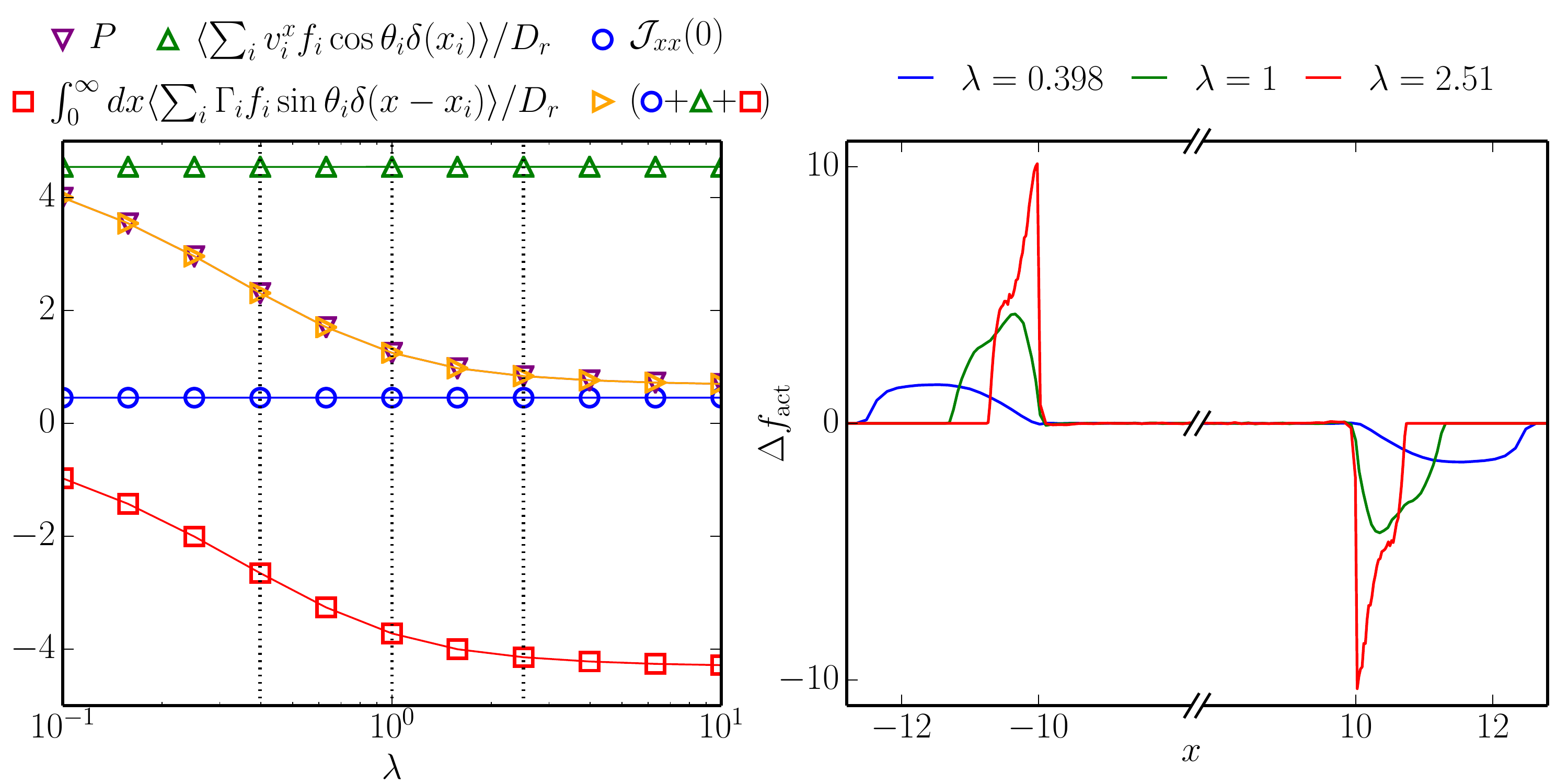}
\caption{Brownian dynamics simulations of self-propelled ellipses with bulk density $\rho_0=1$ confined between harmonic walls located at $x=\pm x_w$ for $x_w=10$, $f_i=1$, $D_r=0.1$, $D_t=0$, $\tilde \gamma=1$, $m=1$, $\kappa=1$.  {\bf
    Left:}  Wall pressure and its various contributions as a function of the wall stiffness $\lambda$. The pressure measured from the definition Eq.~\eqref{eq:globalpressure} (purple triangles) matches the one obtained from Eq.~\eqref{eq:pressureEOStorque} (yellow triangles). Of the three contributions to the latter (green triangles, blue circles, red squares), only the one due to wall torques (red squares) depends on the wall stiffness, leading to a breakdown of the EOS.  {\bf Right:} Spatial profiles of the wall-dependent momentum sources defined by Eq.~\eqref{eq:source} for three different values of $\lambda$ showing its localization in the vicinity of the walls. The three values of $\lambda$ appear as vertical dotted lines in the left panel.  }\label{sourcesinks}
\end{figure}

The three contributions to the pressure $P$ defined in
Eq.~\eqref{eq:pressureEOStorque} are shown in the left panel of
Fig.~\ref{sourcesinks} for walls with
$\lambda_R=\lambda_L=\lambda$. The figure shows that only the
torque-dependent contributions depend on the wall stiffness
$\lambda$. The corresponding wall-dependent sources and sinks $\Delta f_{\rm
  act}(x)$ are shown in the right panel of Fig.~\ref{sourcesinks} for
three stiffnesses.

\begin{figure}[h]
\includegraphics[width=\textwidth]{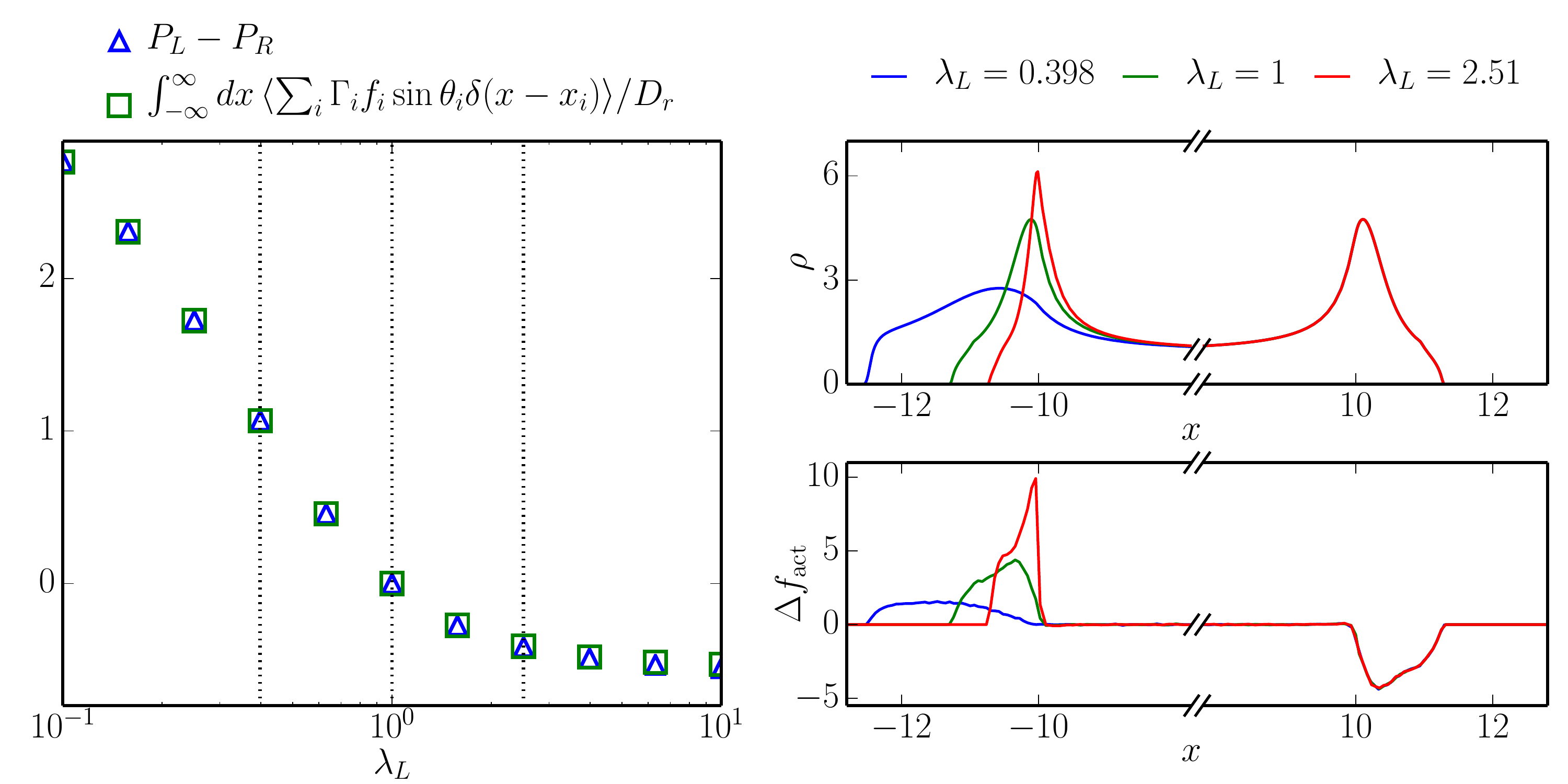}
\caption{Brownian dynamics simulations of self-propelled ellipses
  confined between harmonic walls with stiffnesses $\lambda_R=1$ on
  the right and $\lambda_L$ on the left. All other parameters are
  identical to Fig.~\ref{sourcesinks}.  {\bf Left:} Pressure
  difference between the two walls as a function of
  $\lambda_L$. Different stiffnesses ($\lambda_L\neq\lambda_R$) lead
  to a difference of pressure between the two walls (blue
  triangles). It is fully accounted for by the torque-dependent
  contribution to the total active force (green squares).  {\bf
    Right:} Spatial profile of the density (top) and the
  wall-dependent momentum source defined by Eq.~\eqref{eq:source} (bottom) for
  three different values of $\lambda_L$, indicated by vertical dotted
  lines in the left panel. }\label{fig:totalforce}
\end{figure}

An important consequence of those sources appears when the particles
are confined by left and right walls with different stiffness
$\lambda_R\neq \lambda_L$. Integrating Eq.~\eqref{eq:localpressure}
over the whole space shows that the difference between the pressures
measured on the left and right walls is equal to the total active
force in the system:
\begin{equation}
  P_R-P_L\equiv\int_{-\infty}^\infty \sum_i \langle f_i \cos\theta_i\delta(x-x_i) \rangle dx
\end{equation}
This shows that when the system is confined between two walls with
different stiffnesses there may be a net force acting on the
boundaries. Eq.~\eqref{eq:forcebalance} then shows this total active
force to be given by the torque-dependent sources
\begin{equation}
  P_R-P_L=\int_{-\infty}^\infty \Delta f_{\rm act}(x) dx  = - \int_{-\infty}^\infty \sum_i \langle \frac{f_i}{D_r} \Gamma_i \sin\theta_i \delta(x-x_i) \rangle dx\;,
\end{equation}
as illustrated in Fig.~\ref{fig:totalforce}. This validates our
interpretation of $\Delta f_{\rm act}$ as the net steady-state sources
and sinks of momentum. So far, we have shown that the existence of an
EOS depends on the fact that the dynamics of the momentum density
field takes the form of a conservation equation in steady-state. We
now discuss in more detail the underlying physical interpretation.

\subsection{Torque-free active gas:  active impuse and the emergence of an equation of state}
\label{sec:impulse}

We now consider torque-free active particles and discuss why an
equation of state is recovered in this case. In particular, we show
this to be a property shared by all systems where the dynamics of the
active force $f_i {\bf u_i}$ are decoupled from other degrees of
freedom and lead to $\langle f_i {\bf u_i} \rangle=0$ in
steady-state.  In the main text, we consider the case of torque-free rotational
diffusion while a more general case is addressed in~\ref{sec:SCEOS}. 

The contribution of the activity to the pressure stems from the
momentum transferred to the particles through the active forces. This
can be quantified by the ``active impulse'' which measures the
momentum the active particle will receive on average from its active
force in the future.  For the case of pure, torque-free rotational
diffusion, this can be readily computed as
\begin{equation}\label{eq:activeimpulse}
\Delta {\bf p}_i^{\rm a}(t)\equiv \int_t^\infty \overline{f_i {\bf
u[\theta_i(s)]}} ds = \frac{f_i}{D_r} {\bf u}[\theta_i(t)]
\end{equation}
where the overline denotes an average with respect to histories of the
system in the time interval $[t,+\infty)$ for a given value of
  $\theta_i(t)$.  In~\eqref{eq:activeimpulse}, the active impulse
  simply depends on the initial angle $\theta_i(t)$ because the
  dynamics of the active force $f {\bf u_i}$ is independent of all
  other degrees of freedom. For a more general discussion, we refer
  the reader to~\ref{sec:SCEOS}.

By construction, the dynamics of the active impulse
  obey
\begin{equation}\label{eq:activeimpulsedyn}
\partial_t \Delta {\bf p}_i^{\rm a}(t)= - f_i {\bf u}[\theta_i(t)] 
 \;.
\end{equation}  
In turn, the dynamics of the mean active-impulse field $\langle \Delta
{\bf p^{\rm a}}(x)\rangle = \langle \sum_i\Delta {\bf p}_i^a
\delta({\bf r}-{\bf r_i}) \rangle$ is given by:
\begin{equation}\label{eq:dynFID}
\partial_t \langle \Delta {\bf p^{\rm a}}(x) \rangle= - \langle \sum_i f_i {\bf
u}[\theta_i(s)] \delta({\bf r}-{\bf r_i}) \rangle- \nabla \cdot \langle \sum_i
{\bf v_i} \Delta{\bf  p^{\rm a}_i} \delta({\bf r}-{\bf r_i}) \rangle
\end{equation}
where the divergence $\nabla$ is contracted with the velocities ${\bf
  v_i}$. This gives in the steady state
\begin{eqnarray}
\langle \sum_i f_i {\bf u}[\theta_i(s)] \delta({\bf r}-{\bf r_i}) \rangle &=& -
\nabla \cdot \langle \sum_i
{\bf v_i} \Delta{\bf p^{\rm a}_i} \delta({\bf r}-{\bf r_i}) \rangle\\&=& -
\nabla \cdot \langle \sum_i
{\bf v_i} \frac{f_i}{D_r}{\bf u }(\theta_i) \delta({\bf r}-{\bf r_i}) \rangle \;.\label{activeforcediv}
\end{eqnarray}

Despite each active force injecting momentum into the system,
Eq.~\eqref{activeforcediv} shows that their average contribution in
the steady state takes a momentum-conserving form, namely the mean
local active force can be written as the divergence of a local tensor.
This can be understood as follows. In any volume element, the mean
active force decays to zero because of rotational diffusion. A
non-vanishing mean local active force can thus only be sustained by
incoming fluxes of particles which carry their active force with
them. This is quantified by the tensor measuring the flux of active
impulse
\begin{equation}\label{eq:activeimpulsetensor} {\cal G}\equiv
  \sum_i {\bf v_i} \frac{f_i}{D_r}{\bf u}(\theta_i)  \delta({\bf
    r}-{\bf r_i})\;.
\end{equation}
$\langle {\cal G}({\bf r}) \rangle$ is non-zero because of the
correlations between the velocities of the particles and their active
forces.

\begin{figure}[h]
  \begin{center}
  \includegraphics{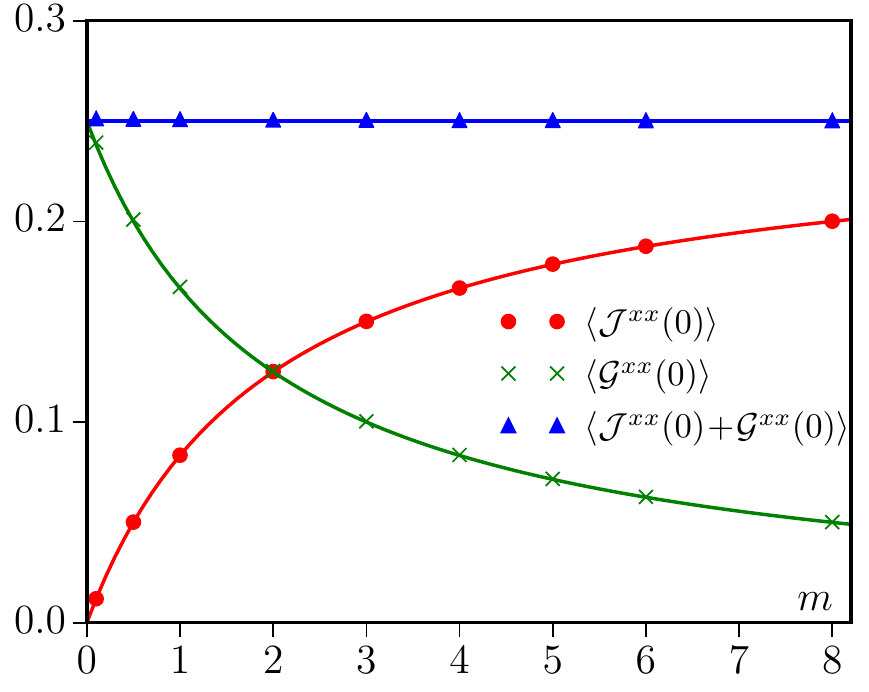}
\end{center}
\caption{Brownian dynamics simulation of self-propelled disks with
  $\tilde\gamma=2$, $f_i=1$, $D_r=1$, $\rho_0=1$, $\lambda=1$,
  $D_t=0$. The mean momentum and active impulse fluxes $\langle {\cal
    J}^{xx}(0)\rangle $ and $\langle {\cal G}^{xx}(0) \rangle$ depend
  on the particles mass $m$ but their sum, which yields the pressure,
  remains constant. Solid lines correspond to the theoretical
  predictions~\eqref{eq:GJtheory} and~\eqref{eq:pressureresult} while
  symbols stem from numerical measurements.}
\label{fig:pressurevsm}
\end{figure}

With the above results, expression~\eqref{eq:globalpressure} for the
pressure can now be written as
\begin{equation}\label{eq:pressureEOS}
  P =\langle {\cal J}^{xx}({x_b}) \rangle 
+  \int_{x_b}^{\infty} \langle\sum_i  f_i \cos \theta_i \delta(x-x_i) \rangle dx=
\langle {\cal J}^{xx}({x_b}) \rangle  + \langle {\cal G}^{xx}({x_b})\rangle.
\end{equation}
This expression clearly depends solely on bulk quantities and
constitutes the EOS of the mechanical pressure in the absence of
torques. The correlators appearing in~\eqref{eq:pressureEOS} can be
computed using It\=o-calculus. Let us first compute
\begin{eqnarray}
  \partial_t \langle \sum_i f_i \cos\theta_i v_i^x \delta(x-x_i) \rangle &=&-\frac{\tilde \gamma}{m} \langle \sum_i f_i\cos\theta_i v_i^x \delta(x-x_i) \rangle +\langle \sum_i \frac 1 m f_i^2\cos^2\theta_i \delta(x-x_i)\rangle\nonumber\\&& -D_r\langle\sum_i f_i\cos\theta_i v_i^x \delta(x-x_i) \rangle - \partial_x \langle \sum_i f_i \cos\theta_i (v_i^x)^2 \delta(x-x_i) \rangle\nonumber
\end{eqnarray}
Since all particles have the same propulsive force $f_i=f$, one finds
in a homogeneous, isotropic bulk of density $\rho_0$, where $\langle
\cos^2\theta_i\rangle = 1/2$, that
\begin{equation}
  \big(D_r+\frac{\tilde\gamma}{m}\big) \langle \sum_i f\cos\theta_i v_i^x \delta(x-x_i) \rangle =  \frac{\rho_0 f^2}{2 m}\label{tensor1}
\end{equation}
We then consider
\begin{eqnarray}
  \partial_t \langle \sum_i \frac {m(v_i^x)^2} 2  \delta(x-x_i) \rangle &=& -\tilde \gamma \langle \sum_i (v_i^x)^2\delta(x-x_i) \rangle + \langle \sum_i f_i v_i^x \cos\theta_i \delta(x-x_i)\rangle\nonumber \\&& + \frac {\tilde{\gamma}^2 D_t}m\sum_i\delta(x-x_i)-\partial_x \langle \sum_i \frac {m(v_i^x)^3} 2  \delta(x-x_i) \rangle\nonumber
\end{eqnarray}
which yields in the steady-state of the homogeneous bulk
\begin{equation}\label{tensor2}
  \langle \sum_i (v_i^x)^2\delta(x-x_i) \rangle = \frac{\tilde \gamma \rho_0 D_t}m + \frac 1 {\tilde \gamma}  \langle \sum_i f_i v_i^x \cos\theta_i \delta(x-x_i)\rangle
\end{equation}
Using both the definitions~\eqref{eq:defJ}
and~\eqref{eq:activeimpulsetensor} and the results~\eqref{tensor1}
and~\eqref{tensor2}, one then finds
\begin{equation}\label{eq:GJtheory}
  \langle {\cal G}^{xx}({x_b})  \rangle = \rho_0 \frac{f^2}{2 D_r (m D_r + \tilde \gamma)}\qquad\mbox{and}\qquad \langle {\cal J}^{xx}({x_b})\rangle = \rho_0 \Big(\tilde{\gamma} D_t+\frac{m f^2}{2\tilde\gamma (\tilde \gamma+mD_r)}\Big)
\end{equation}
leading to the previously known expression for the pressure of a
torque-less dry active system
\begin{equation}\label{eq:pressureresult}
  P=\rho_0 \left (\tilde{\gamma} D_t + \frac{f^2}{2 D_r \tilde \gamma} \right )
\end{equation}
Despite $ \langle {\cal G}^{xx}({x_b}) \rangle$ and $\langle {\cal
  J}^{xx}({x_b})\rangle$ both depending on the mass of the particles,
Eq.~\eqref{eq:pressureresult} shows the overall pressure to be
independant of the mass. (This is illustrated in
Fig.~\ref{fig:pressurevsm} for the system introduced in
section~\ref{sec:numericstorques}.) Overdamped and underdamped dynamics
thus have the same pressure.

It is illuminating to restate the above discussion for the appearance
of an equation of state from both a global and a local picture.
\paragraph{Global picture of the EOS.}
Eq.~\eqref{eq:pressureEOS} relies on the fact that the total active
force exerted in the $x>{x_b}$ region is equal to the flux of free active
impulse through the $x={x_b}$ plane. In the absence of any bias in the
dynamics of the active force, an active particle experiences on average
a vanishing active force in steady-state. The presence of a wall at the
right end of the system may alter its trajectory, but not the
statistics of its active force. The only contribution that makes the
mean active force non-zero in the $x>{x_b}$ region is thus that particles
entering (leaving) this region typically have a positive
(negative) component of their propulsive force along $x$. The
magnitude of the mean active force in this region is thus directly measured
by the flux of active impulse through the $x={x_b}$ interface.

\paragraph{Local picture of the EOS.}
Let us now discuss what happens in the vicinity of a confining wall at
the right end of the system. In the bulk, the flux of momentum $\langle {\cal
  J}^{xx} \rangle$ and of active impulse $\langle {\cal G}^{xx} \rangle$ are uniform. In the
presence of an external force, Eqs.~\eqref{eq:localpressure}
and~(\ref{activeforcediv}-\ref{eq:activeimpulsetensor}) lead to
\begin{equation}\label{eqn:forcedensitybalance}
\rho \partial_x V_{\rm ext} =  - \partial_x\left[  \langle {\cal J}^{xx}+{\cal G}^{xx} \rangle\right]\,.
\end{equation}
This states that the force (density) exerted by the wall generates
a decay of the incoming fluxes of momentum and active impulse. This
can be easily understood: for a passive system, the wall stops the
particles by applying forces that decrease their momentum. The overall
drop of the momentum flux from its bulk value to zero is then nothing
but the pressure. For active particles, in addition to decreasing the
momentum of the active particles, the wall has to compensate the
momentum they  gain due to the active force, i.e. it has to
``consume'' the active impulse of the particles. 

\begin{figure}
  \begin{center}
    \includegraphics[width=.8\textwidth]{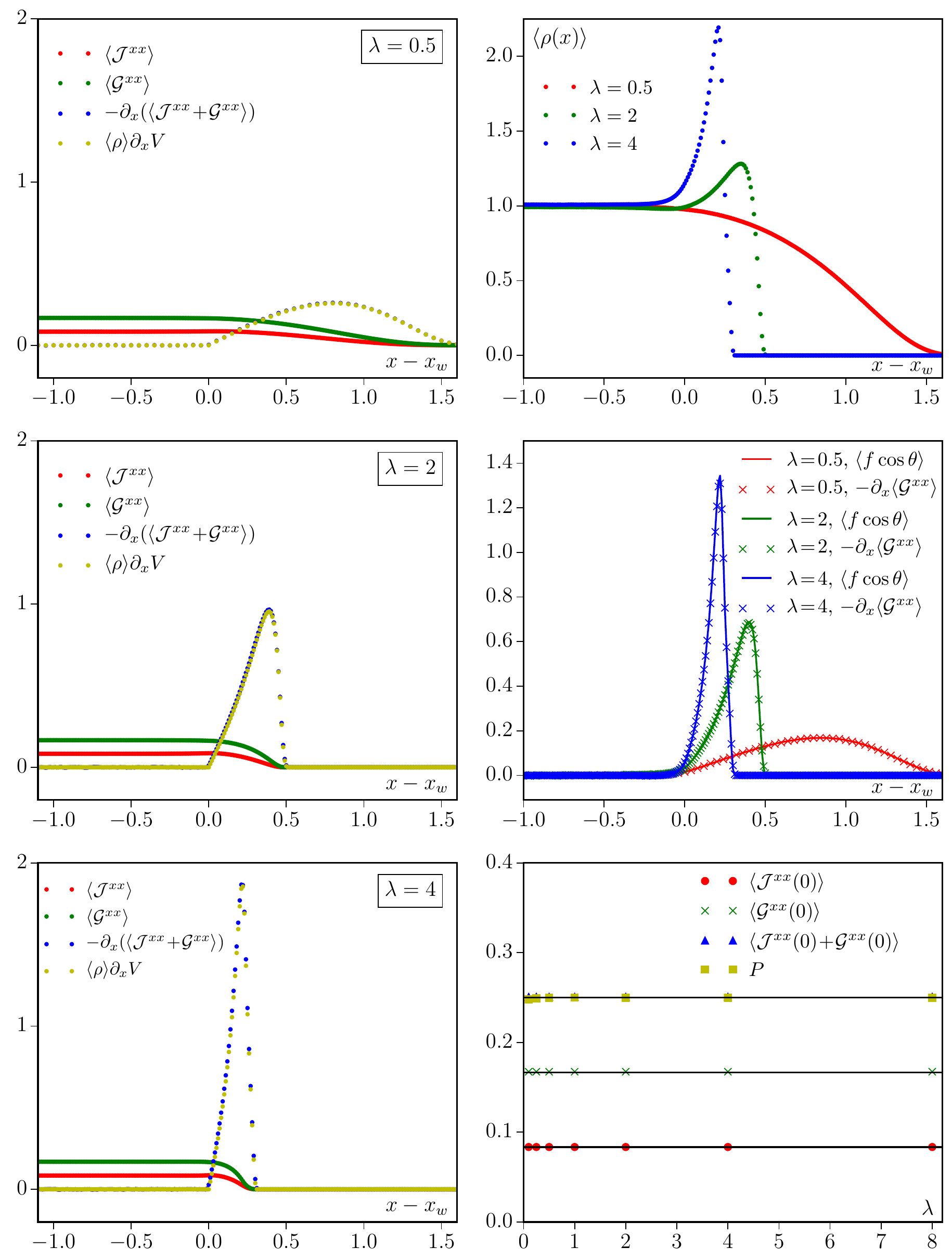}
  \end{center}
  \caption{Brownian dynamics simulations of self-propelled disks
    confined by walls of different stiffness ($\lambda=0.5,\,2,\,4$)
    for {$f_i=1$, $D_r=1$, $D_t=0$, $\tilde \gamma=2$, $m=1$,
      $\rho_0=1$}. {\bf Left:} Momentum (red) and active impulse (green)
    fluxes. The wall force (yellow) is exactly balanced by the
    divergence of the incoming momentum and active impulse
    fluxes. {\bf Top right}: Number density showing wall-dependent
    boundary layers. {\bf Middle right:} The average active force is
    given by (minus) the divergence of the incoming active impulse
    flux. {\bf Bottom right:} Numerical measurements of the pressure,
    $\langle {\cal G}^{xx}(0)\rangle $, $\langle{\cal
      J}^{xx}(0)\rangle$ and their sum (symbols), compared with their
    theoretical predictions (solid lines) for varying wall
    stiffness.}\label{fig:EOSandfluxes}
\end{figure}

Equation~\eqref{eqn:forcedensitybalance} shows the wall force to
generate a non-zero divergence of the momentum and active impulse
fluxes. In turn, the latter yield a layer of active force density
close to the wall, according to
Eqs.~(\ref{activeforcediv}-\ref{eq:activeimpulsetensor}). The shape of
this layer will typically depend on the confining potential, but since
the integrated action of the wall potential is to bring the incoming
fluxes from their bulk value to zero, the total active force, and
hence its contribution to pressure, will not depend on the wall
potential. This is illustrated in Fig.~\eqref{fig:EOSandfluxes} for
the system described in section~\ref{sec:numericstorques}.

\subsection{Active impulse and swim pressure}
\label{app:swimpressure}
Before we discuss in more detail the connection between the EOS and
the effective conservation of momentum, we note that the flux of
active impulse $\langle {\cal G} \rangle$ can be expressed in a
slightly different manner. To do so, we use
\begin{equation}
  \partial_t  \langle {\bf r}_i  f_i {\bf u}(\theta_i) \rangle = \langle {\bf v}_i  f_i {\bf u}(\theta_i) \rangle - D_r \langle {\bf r}_i f_i {\bf u}(\theta_i) \rangle \;,
\end{equation}
Therefore, in the steady state one can express $\langle {\cal G}
\rangle$ as
\begin{equation}\label{eq:swimpressure}
  \langle {\cal G}(\bfr) \rangle = \langle \delta({\bf r}-{\bf r_i}) {\bf r}_i f_i {\bf u}(\theta_i) \rangle \;.
\end{equation}
This expression was introduced in \cite{Takatori2014} and termed the
swim-pressure. It can also be found using an approach generalizing the
virial theorem~\cite{Winkler2015SoftMatter,Falasco:2016:NJP} or
following Irving and Kirkwood~\cite{Yang2014}. Note, however, that the
discussion above makes it clear that this does not, in general,
provide the expression for the pressure unless an effective momentum
conservation within the active fluid is present in the steady
state. Moreover, the equality between the active force and ${\cal G}$,
whether its expressions~\eqref{eq:activeimpulsetensor}
or~\eqref{eq:swimpressure}, only holds in the steady state. {Last,
  conservation laws are typically defined dynamically, whereas our
  derivation only holds in the steady-state, we now discuss a more
  precise definition of what we mean by effective momentum
  conservation in the steady-state.}

\subsection{Effective steady-state momentum conservation}
\label{app:momcon}
Let us first consider the torque-free non-interacting active particles
of section~\ref{sec:impulse}. In this case, the dynamics of the mean
momentum density field can be obtained from Eq.~\eqref{mom_dynamics}
\begin{equation}
  \partial_t \vec{p} = -\gamma \vec{ p} -   \rho
\nabla V_{\rm ext}+\langle \sum_i  f_i \vec
u(\theta_i) \delta(\vec r-\vec r_i) \rangle      -
\grad\cdot [  \langle {\cal J} \rangle ] \;.
\end{equation}
There are three  sources of momentum for the active particles in
the steady state: the external wall, through $-\rho\grad V_{\rm ext}$,
the friction with the substrate, through $-\gamma \bfp$, and the
active forces. Using the definition of the active
impulse~\eqref{eq:activeimpulse} and its dynamics~\eqref{eq:dynFID},
the mean active force density can always be replaced by
\begin{equation}
\langle \sum_i f_i {\bf
u}[\theta_i(s)] \delta({\bf r}-{\bf r_i}) \rangle=-\partial_t \langle \Delta {\bf p^{\rm a}}(x) \rangle- \nabla \cdot \langle \sum_i
{\bf v_i} \Delta{\bf  p^{\rm a}_i} \delta({\bf r}-{\bf r_i}) \rangle
\end{equation}
so that the dynamics of the momentum field can be written as
\begin{equation}\label{eq:momconsSS}
  \partial_t \vec{p}  = -\gamma \vec{ p} -\partial_t \langle \Delta {\bf p^{\rm a}}(x) \rangle -   \rho
\nabla V_{\rm ext}   -
\grad\cdot [  \langle {\cal J} + {\cal G} \rangle ] \;.
\end{equation}
In steady-state, the first two terms on the right-hand side vanish and
the sole non-vanishing momentum source is  the external
wall. Outside steady-state, however, $\bfp$ is not conserved even in
the absence of external walls. Note that Eq.~\eqref{eq:momconsSS} can
be rewritten as
\begin{equation}\label{eq:momconsSS2}
  \partial_t [\vec{p}+\langle \Delta {\bf p^{\rm a}}(x) \rangle]  = -\gamma \vec{ p}   -   \rho
  \nabla V_{\rm ext}   -
  \grad\cdot [  \langle {\cal J} + {\cal G} \rangle ] \;.
\end{equation}
This equation states that, apart from the friction force and the
external forces, the sum of the momentum of the particles $\bfp$ and
of their active impulse $\bfdpf$ is conserved. The latter acts as a
momentum reservoir for the particles: when the active impulse varies, it is through a
transfer of momentum to the active particles via the active
forces. Since the friction term $-\gamma \bfp$ vanishes in
steady-state, this is probably the cleanest way of expressing what we
mean by effective steady-state conservation of momentum in this
system.

The breakdown of the equation of state in the case with external
torques can then be understood from two different viewpoints. First,
one can keep the definition of the active impulse of a particle as an
intrinsic property:
\begin{equation}\label{eq:localdefai}
  \Delta \bfp_i^{\rm a}(t)\equiv \frac{f_i}{D_r} \bfu(\theta_i(t))
\end{equation}
which corresponds to the active impulse of a particle in the absence
of any external wall (say, for periodic boundary conditions). In the
presence of wall torques, Eq.~\eqref{eq:momconsSS2} then becomes
\begin{equation}\label{eq:momconsSS3}
  \partial_t [\vec{p}+\langle \Delta {\bf p^{\rm a}}(x) \rangle]  = -\gamma \vec{ p}   -   \rho
  \nabla V_{\rm ext} -\int_{x_b}^\infty dx \langle \frac{f_i}{D_r} \Gamma_i \sin \theta_i \delta (x-x_i) \rangle
  -
  \grad\cdot [  \langle {\cal J} + {\cal G} \rangle ] \;.
\end{equation}
This highlights a new source of momentum for the particles stemming
from wall torques. 

An alternative view can be obtained if one keeps the definition of
the active impulse as
\begin{equation}\label{eq:defainl}
\Delta {\bf p}_i^{\rm a}(t)\equiv \int_t^\infty \overline{f_i {\bf
u[\theta_i(s)]}} ds\,,\quad\mbox{whence}\quad \bfdp(t) \neq \frac{f_i}{D_r} {\bf u}[\theta_i(t)]\,.
\end{equation}
The last equality does not hold since encounters between a particle
and the wall changes its orientation and hence its momentum and active
impulse: the active impulse is then not a local quantity anymore. This
can be understood by noting that the definition~\eqref{eq:defainl} is
non-local in time. This translates into a non-locality in space
because of the motion of the particles during $[t,+\infty)$ and the
  active impulse of a particle in the bulk of the system depends on
  its fate upon future encounters with the walls. With this
  definition, one can then still relate the pressure to the fluxes of
  momentum and of active impulse through
\begin{equation}\label{eq:newimp}
  P = \langle {\cal J}^{xx}+{\cal G}^{xx}\rangle\quad\mbox{with}\quad {\cal G}\equiv \sum_i \bfv_i \bfdp \delta(\bfr-\bfr_i)\,.
\end{equation}
But ${\cal G}$ is not given anymore by the local
tensor~\eqref{eq:activeimpulsetensor} and becomes a \textit{non-local}
quantity, which depends on the wall stiffness so
that~\eqref{eq:newimp} does not yield an equation of state
anymore. Both interpretations are equally valid,
but~\eqref{eq:localdefai} makes the role of wall torques as momentum
sources more explicit and this is the path we follow in this article.

Now that we have extensively discussed the case of non-interacting
active particles, we show our results to extend to interacting active
particles in section~\ref{sec:interactions} and discuss their
implications for phase-separating systems in section~\ref{sec:MIPS}.

\section{Interacting active particles}

\label{sec:interactions}

In what follows we now include interactions between particles and
study their effect on the existence or absence of an equation of
state. To do this, we consider torque-free particles and look at two
models---quorum sensing interactions and pairwise interactions. In the
overdamped limit, it is known that the first does not admit an equation
of state while the latter
does~\cite{Solon:2015:NatPhys,Solon_interactions}. As shown below this
also holds for the underdamped model and is, as in the non-interacting
case, directly related to the presence or absence of momentum sources
and sinks in the steady state.

The equations of motion, allowing for both quorum sensing and pairwise
interactions, are:
\begin{eqnarray}
\label{eq:manybod_eqs}
\dot{{\bf r}}_i&=&{\bf v}_i \nonumber \\
m\dot{{\bf v}}_i&=&-\tilde{\gamma} {\bf v}_i +f_i(\lbrace {\bf r} \rbrace) {\bf u}(\theta_i)+{\bf F}_i(\lbrace {\bf r} \rbrace)-\nabla_{{\bf r}_i} V_{ext} +\sqrt{2 \tilde{\gamma}^2 D_t}{\bf \eta}_i \nonumber \\
\dot{\theta}_i&=& \sqrt{2D_r} \zeta_i \;,
\end{eqnarray}
where $ \lbrace {\bf r} \rbrace=\lbrace {\bf r}_1,{\bf r_2}, \ldots
\rbrace$ is the collection of coordinates of all the particles. Note
that we allow the magnitude of the active force $f_i(\lbrace {\bf r}
\rbrace)$ to depend on the locations of the other particles. In
addition we consider inter-particle forces ${\bf F}_i(\lbrace {\bf r}
\rbrace)$. All these interactions are assumed to be short ranged. It
is a straightforward exercise to extend the model to include a
dependence of the active force $f_i$ and the force ${\bf F}_i$ on the
orientations of all the particles or to include interparticle aligning
torques (see~\cite{Solon:2015:NatPhys} for a discussion of this case
in the overdamped limit). Since such angular dependences do not alter
our conclusions qualitatively, we do not detail here the results
concerning these cases.

As before,  we consider the dynamics of the momentum
density field, which is now given by
\begin{equation}
\label{eq:manybod_mom_dynamics}
\partial_t {\bf p}= -\gamma {\bf p} -  \rho \nabla V_{ext}  +\sum_i \langle f_i(\lbrace {\bf r} \rbrace) {\bf u}(\theta_i) \delta({\bf r}-{\bf r}_i) \rangle  + \sum_i \langle {\bf F}_i(\lbrace {\bf r} \rbrace) \delta ({\bf r}-{\bf r}_i) \rangle - \nabla \cdot \langle {\cal J} \rangle 
\end{equation}
where we use the notations of the previous section.
For the flat wall geometry of Fig.~\ref{fig:walls}, one gets
\begin{equation}\label{eq:manybod_globalpressure}
  P =  \langle {\cal J}^{xx}({x_b}) \rangle 
+  \int_{x_b}^{\infty} \langle\sum_i  f_i(\lbrace {\bf r} \rbrace) \cos \theta_i \delta(x-x_i) \rangle dx + \int_{x_b}^\infty \langle \sum_i {F^x}_i(\lbrace {\bf r} \rbrace) \delta (x-x_i) \rangle dx\;,
\end{equation}
where $F^x_i$ is the $x$--component of the force ${\bf F}_i$ and we use
the convention introduced before Eq.~\eqref{eq:localpressure} so that an
integration over the $\hat y$ direction is carried out. To look for a
possible effective momentum conservation which, much like in the
non-interacting case, will lead to an equation of state, one needs to
analyze the last two terms in
Eq.~\eqref{eq:manybod_globalpressure}. Note that, as before, we assume
that the bulk steady-state is uniform.

In what follows it will be helpful to use the relation 
\begin{eqnarray}
\partial_t \langle f_i \cos \theta_i \delta (x-x_i) \rangle &=&  \sum_j \langle {\bf v}_j \cdot (\nabla_{{\bf r}_j} f_i) \cos \theta_i \delta (x-x_i) \rangle +\langle v^x_i f_i \cos \theta_i \partial_{x_i} \delta (x-x_i) \rangle \nonumber \\ &-&D_r \langle  f_i \cos \theta_i \delta (x-x_i)  \rangle \;,
\end{eqnarray}
which  gives in the steady state
\begin{eqnarray}
\label{eq:manybod_cos}
	 \langle  f_i \cos \theta_i \delta (x-x_i)  \rangle&=& \sum_j \langle {\bf v}_j \cdot (\nabla_{{\bf r}_j} \frac{f_i}{D_r}) \cos \theta_i \delta (x-x_i) \rangle - \partial_x \langle v^x_i \frac{f_i}{D_r} \cos \theta_i \delta (x-x_i) \rangle\\
&=& \sum_j \langle {\bf v}_j \cdot (\nabla_{{\bf r}_j} \frac{f_i}{D_r}) \cos \theta_i \delta (x-x_i) \rangle - \partial_x \langle {\cal G}^{xx} \rangle\,.
\end{eqnarray}
This shows that, due to the dependence of the self-propulsion force
$f_i$ on the ${\bf r}_j$'s, the density of active forces is not purely
given by the divergence of the flux of the active impulse ${\cal G}$
defined in~\eqref{eq:activeimpulsetensor}. Together with~Eq.~\eqref{eq:manybod_globalpressure}, this suggests a potential
breakdown of the equation of state. The measure of active sources
$\Delta f_{\rm act}$, defined in Eq.~\eqref{eq:source}, is now given by
\begin{eqnarray}\label{eq:sourceQS}
\Delta f_{\rm act}(x)&=&  \langle \sum_i f_i \cos \theta_i \delta(x-x_i) \rangle +\partial_x \langle \frac{v_i^x}{D_r} f_i \cos\theta_i \delta(x-x_i)\rangle
  \rangle\\
  &=&\sum_j \langle {\bf v}_j \cdot (\nabla_{{\bf r}_j} \frac{f_i}{D_r}) \cos \theta_i \delta (x-x_i) \rangle
\end{eqnarray}
Let us first treat  the case of pure quorum sensing and discuss
afterwards interparticle forces ${\bf F}_i(\lbrace {\bf r} \rbrace)$.

\subsection{Quorum sensing}

In models of quorum sensing the active force of a particle is
modulated according to the density of particles in a region around
it. To make the discussion clearer we assume that ${\bf F}_i(\lbrace
{\bf r} \rbrace)=0$. Using Eq.~\eqref{eq:manybod_cos} one finds that the
pressure is given by
\begin{eqnarray}
  P &=&  \langle {\cal J}^{xx}({x_b}) \rangle + \langle {\cal G}^{xx}({x_b}) \rangle
+  \int_{x_b}^{\infty} \langle \sum_j {\bf v}_j \cdot (\nabla_{{\bf r}_j}) \frac{f_i}{D_r}\cos \theta_i \delta (x-x_i) \rangle dx\\
 &=&  \langle {\cal J}^{xx}({x_b}) \rangle + \langle {\cal G}^{xx}({x_b}) \rangle
+  \int_{x_b}^{\infty} \Delta f_{\rm act}(x) dx\label{eq:manybod_globalpressure_quo}
\end{eqnarray}
where as before ${x_b}$ is a point in the bulk of the system. The only
term which may act as a momentum source/sink is the last one. At any
position $x_{\rm b}$ in the bulk, the system is isotropic in the
steady-state and
\begin{equation}
\Delta f_{\rm act}(x_b)=  \langle \sum_j {\bf v}_j \cdot (\nabla_{{\bf r}_j}) \frac{f_i}{D_r}\cos \theta_i \delta (x_{\rm b}-x_i) \rangle=0
\end{equation}
because of the angular average. This, however, is not true in the
presence of an external potential and hence near the wall. As we now
show, this leads to the absence of an equation of state. We consider a
model for quorum sensing with
\begin{eqnarray}\label{eq:QSint}
  f_i(\lbrace {\bf r} \rbrace)&=& f_0+\frac{f_1-f_0}{2} \Big[ \tanh\Big(\frac{\tilde \rho({\bf r}_i)-\rho_m}{L_f}\Big)+1\Big]\\
  \tilde \rho({\bf r})&=&\int d{\bf y} K(|{\bf r}-{\bf y}|)\hat\rho({\bf y});\qquad K(r)=Z^{-1} \exp\Big[-\frac{1}{1-r^2}\Big] \Theta(1-r)
\end{eqnarray}
Here, $\tilde \rho({\bf r}_i)$ is a measure of the particle density
$\hat \rho(\bfr_i)$ using a coarse-graining kernel $K$; $Z$ is a normalisation
constant such that $\int d {\bf r} K(|\bfr|)=1$. The constants $f_0$,
$f_1$, $\rho_m$ and $L_f$ control the dependence of $f_i$ on the local
density.

\begin{figure}
  \includegraphics[]{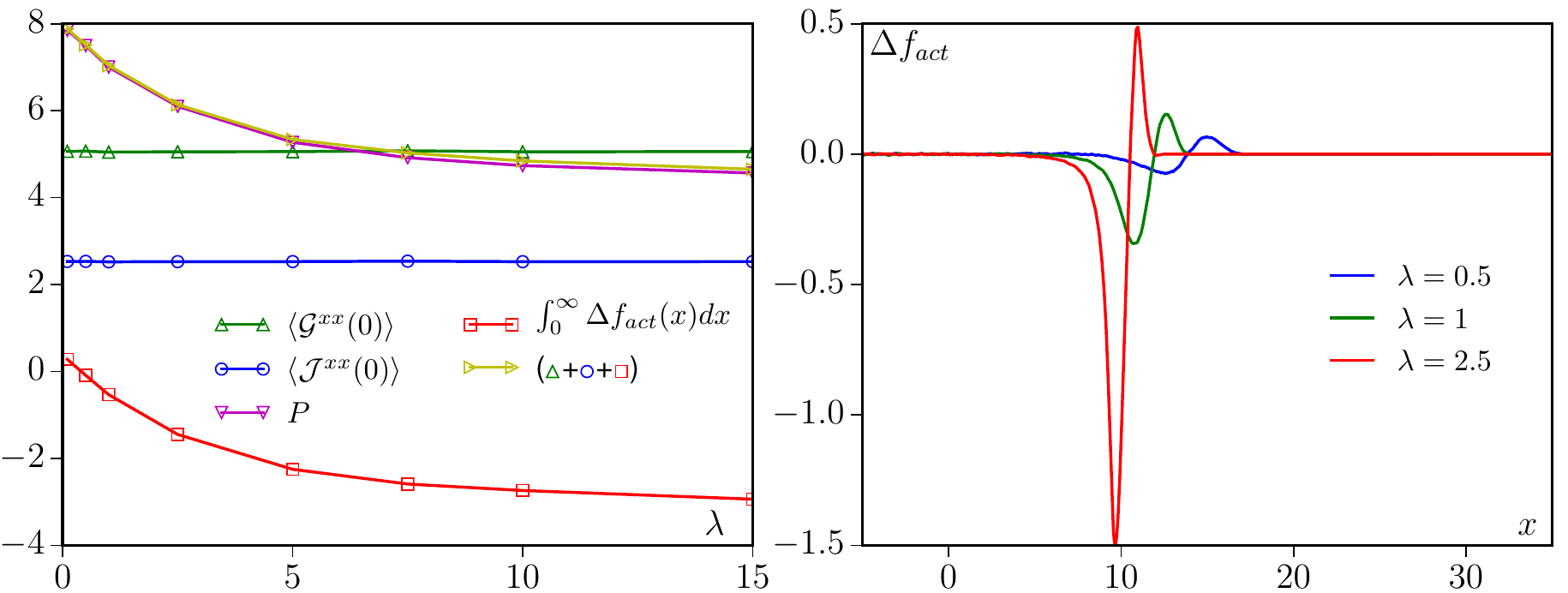}
  \caption{Brownian dynamics simulations of self-propelled disks
    interacting through Eq.~\eqref{eq:QSint} with $f_0=4$, $f_1=1$,
    $\rho_m=2.5$, $L_f=1.25$, $D_r=1$, $\tilde \gamma=1$, $m=1$,
    $dt=5.\,10^{-4}$, $D_t=0$, $\rho_0=0.64$,
    $x_w=10$. {\bf Left}: The pressure and its different contributions
    are shown for various wall stiffness. The momentum source $\Delta
    f_{\rm act}$ is shown to account for the entire wall dependence of
    the pressure. {\bf Right}: Steady-state momentum sources $\Delta f_{\rm act}(x)$ for three different wall stiffness.}\label{fig:QS}
\end{figure}

Figure~\ref{fig:QS} shows the results of simulations of active
Brownian particles interacting through~\eqref{eq:QSint}.  The left
panel shows the different terms in the equation for the pressure in
Eq.~\eqref{eq:manybod_globalpressure_quo} as a function of the wall
strength $\lambda=\lambda_R=\lambda_L$ for the potential of
Eq.~\eqref{eq:potential}. As can be seen the pressure depends on the
wall potential and $\Delta f_{\rm act}$ entirely accounts for this
dependence. On the right panel we plot $\Delta f_{\rm act}(x)$ which
is non-zero near the walls and depends on the wall potential. As for
wall torques, the terms which do not conserve momentum in steady-state
account for the breakdown of the equation of state.

\subsection{Pairwise forces}

We now turn to the case where $f_i(\lbrace {\bf r} \rbrace)=f_i$ and
consider momentum- and energy-conserving short-range pairwise forces
${\bf F}_i(\lbrace {\bf r} \rbrace)$. This case was already studied
in~\cite{Steffenoni:2016:arxiv}. Our derivation is distinct and we focus
here on the role of momentum fluxes. For constant self-propelling
forces, Eq.~\eqref{eq:manybod_cos} for the local density of active
forces reduces to the non-interacting case~\eqref{activeforcediv}. The
expression for the pressure is then:
\begin{equation}
\label{eq:manybod_pressure_pair}
 P =  \langle {\cal J}^{xx}({x_b}) \rangle + \langle {\cal G}^{xx}({x_b}) \rangle
+  \int_{x_b}^{\infty}  \langle \sum_i F^x_i(\lbrace {\bf r} \rbrace) \delta (x-x_i) \rangle  dx \;.
\end{equation}

By the conserving nature of the pairwise forces, Newton's third law
implies that the last term in the equation measures the forces across
the plane $x={x_b}$. Since the forces are assumed to be short ranged this
quantity is local. Let us now show this explicitly for the case of
conservative pairwise forces ${\bf F}_i(\lbrace {\bf r}
\rbrace)=-\sum_j \nabla_{{\bf r}_i} V({\bf r}_i-{\bf r}_j)$.
Remembering that Eq.~\eqref{eq:manybod_pressure_pair} contains an
average over $y$, the last term can be expressed as
\begin{equation}
   \int_{x_b}^{\infty}\langle \sum_i  F^x_i \delta (x-x_i) \rangle dx  =
-\sum_{x_i>{x_b}}\sum_{x_j<{x_b}} \langle \partial_{x_i}V({\bf r}_i-{\bf r}_j) \rangle -\sum_{x_i>{x_b}}\sum_{x_j>{x_b}} \langle
\partial_{x_i}V({\bf r}_i-{\bf r}_j) \rangle \;,
\end{equation}
where $x_i$ denotes the $x$--coordinate of particle $i$. The second term clearly vanishes so that
\begin{equation}\label{eq:pwforces}
   \int_{x_b}^{\infty} \langle\sum_i  F^x_i \delta (x-x_i) \rangle dx  = 
-\langle\sum_{x_i>{x_b}}\sum_{x_j<{x_b}} \partial_{x_i}V({\bf r}_i-{\bf r}_j)\rangle
\end{equation}
which can be rewritten in terms of the density-density correlator
\begin{equation}
  P_D({x_b})\equiv \int_{x_b}^{\infty} \langle\sum_i  F^x_i \delta (x-x_i) \rangle dx	  = -\int_{w_1>{x_b}}\!\!\! \!\!\!d{\bf w_1} \int_{w_2<{x_b}}\!\!\!\!\!\! d {\bf w_2}\, \langle \hat\rho({\bf w}_1)\hat\rho({\bf w}_2)
\rangle \partial_{w_1} V({\bf w}_1-{\bf w}_2) \;,
\end{equation}
with $w_1$ ($w_2$) the $x$--component of ${\bf w}_1$ (${\bf
  w}_2$). $P_D({x_b})$, as stated above, is the force exerted across the
$x={x_b}$ plane. It is essentially the usual equilibrium contribution from
pairwise forces to the pressure, and was identified in active
particles in the overdamped
regime~\cite{Solon_interactions}. {All in all, this yield for the
  pressure
\begin{equation}
\label{eq:manybod_pressure_pair2}
 P =  \langle {\cal J}^{xx}({x_b}) \rangle + \langle {\cal G}^{xx}({x_b}) \rangle
+  P_D({x_b}) \;.
\end{equation}
For short-range interactions, this is a bulk property of the fluid,
and hence independent of the wall potential. Note
that~\eqref{eq:manybod_pressure_pair2} simply states that the
effectively conserved momentum is transfered along the system by both
the flux of particles, which carry with them their momentum $\bfp_i$
and active impulse $\bfdp$, and by interparticle forces; this overall
momentum is then absorbed by the wall.}  It is straightforward to
extend this derivation to cases with multi-particle conserving forces
or to conserving forces which depend on the particles' relative
orientations.

\section{Momentum sources in motility induced phase separation}
\label{sec:MIPS}

One of the most remarkable features of active particles is their
generic tendency to phase separate even in the presence of purely
repulsive
interactions~\cite{Tailleur2008PRL,Thompson2011JSM,Fily2012PRL,Redner2013PRL,Bialke2013EPL,Stenhammar2013PRL,Buttinoni2013PRL,Wysocki2014EPL,Theurkauff2012PRL,Wittkowski2014NC,Takatori2014,Speck2014PRL,Matas2014PRE,Zottl2014PRL,Suma2014EPL,Solon_interactions,Cates2015MIPS,Redner2016PRL,Marchetti2016}. It
was recently shown that depending on the type of interactions
considered, active particles undergoing MIPS lead to coexisting phases
with either equal or unequal mechanical
pressures~\cite{Solon:arxiv:2016}. It is thus natural to probe the
mechanical equilibrium between coexisting phases, i.e. ask for the
presence or absence of momentum sources at the interface.

\if{More generally, a lot of effort has been recently put into identifying
relations between local quantities, say correlators, which
characterize the coexisting phases~\cite{Solon:arxiv:2016}. These can
be, in principle, used to give partial information on the coexistence
curves (for single component systems we need at least one more
relation). As we show below the relations which stems from considering
momentum fluxes, can depend explicitly on the shape of the interface,
in contrast to the equilibrium counterpart. When they do depend on the
interface properties, clearly they cannot be use to construct a phase
diagram using only bulk properties.}\fi

For simplicity, in what follows, we focus our discussion on a flat
interface between two phases whose direction is normal to the $x$
axis. In this case, the dynamics of the momentum density
field~\eqref{eq:manybod_mom_dynamics} leads in steady state to the
balance equation:
\begin{equation}
\label{eq:balance}
\sum_i \langle f_i( \lbrace {\bf r} \rbrace ) {\bf u}(\theta_i) \delta({\bf r}-{\bf r}_i) \rangle  + \sum_i \langle {\bf F}_i( \lbrace {\bf r} \rbrace ) \delta ({\bf r}-{\bf r}_i) \rangle = \nabla \cdot \left(\langle {\cal J} ({\bf r})\rangle \right)\;.
\end{equation} 
Its projection on the one-dimensional geometry, obtained as in Eq.~\eqref{eq:localpressure}, is given by 
\begin{equation}
\label{eq:xbalance}
\sum_i \langle f_i( \lbrace {\bf r} \rbrace ) \cos(\theta_i) \delta(x-x_i) \rangle  + \sum_i \langle F^x_i( \lbrace {\bf r} \rbrace ) \delta (x-x_i) \rangle = \partial_x \langle {\cal J}^{x x} \rangle
\end{equation} 

As above we consider two representing cases, quorum sensing and
pairwise interactions, separately. Before doing so, it is useful to
consider the equilibrium case where $f_i=0$. For concreteness we
assume, say, a high density phase at $x < 0$ a low density phase at $x
> 0$ with an interface between the two phases in the vicinity of
$x=0$. Integrating Eq.~\eqref{eq:xbalance} from a point $x_\ell < 0$ to
$x_r > 0$, assuming that the distance between $x_\ell$ and $x_r$ is much
larger than both the interaction range and the width of the interface,
gives
\begin{equation}
\label{eq:forcebaleq}
{P_D}(x_\ell)+\langle {\cal J}^{x x}(x_\ell) \rangle = {P_D}(x_r)+\langle {\cal J}^{x x}(x_r) \rangle
\end{equation}
Namely, by integrating the momentum balance equation over the
interface, we obtain the equality of mechanical pressures between the
two phases (which each have different values of $P_D$ and $\langle {\cal J}^{\rm
  xx} \rangle$).

We now turn to the two active models.

\subsection{Pairwise forces}
As we showed before in this case, where $f_i( \lbrace {\bf r} \rbrace
)=f_i$, an equation of state for the pressure exists. In addition, it
is well known that, at coexistence, there is an orientational ordering
of the particles at the interface between the two
phases~\cite{Fily2012PRL,Redner2013PRL}. Therefore, naively, the term
$\sum_i \langle f_i( \lbrace {\bf r} \rbrace ) \cos(\theta_i)
\delta(x-x_i) \rangle $ in Eq.~\eqref{eq:xbalance} might lead one to
conclude that there are steady-state momentum sources in the
system. However, similarly to the ordering at confining walls, this is
not the case. In fact, using Eq.~\eqref{eq:localmeanactiveforce} and
\ref{eq:activeimpulsetensor} we can rewrite Eq.~\eqref{eq:xbalance} as
\begin{equation}
\label{eq:xbalance_pair}
 \sum_i \langle F^x_i( \lbrace {\bf r} \rbrace ) \delta (x-x_i) \rangle = \partial_x \left[\langle {\cal J}^{xx}  + {\cal G}^{xx} \rangle \right] \;.
\end{equation} 
This equation implies that the forces between the particles, which in
equilibrium systems are compensated only by the momentum flux, are
here compensated by both impulse and momentum fluxes. Repeating the
steps leading to Eq.~\eqref{eq:forcebaleq} now gives an equality of
mechanical pressures between the two phases
\begin{equation}
\label{eq:forcebalpair}
P_D(x_\ell)+\langle {\cal J}^{x x}(x_\ell) +{\cal G}^{ xx}(x_\ell)\rangle = P_D(x_r)+\langle {\cal J}^{x x}(x_r)+ {\cal G}^{ xx}(x_r) \rangle \;.
\end{equation}
Despite the presence of a layer of active forces localized at the
interface, the mechanical pressures of coexisting phases are
equal. This highlights again the effective momentum conservation in
this system, as discussed in~\ref{app:momcon}, and the absence of
momentum sources in steay state.

\begin{figure}
  \begin{center}
    \includegraphics[width=0.8\textwidth]{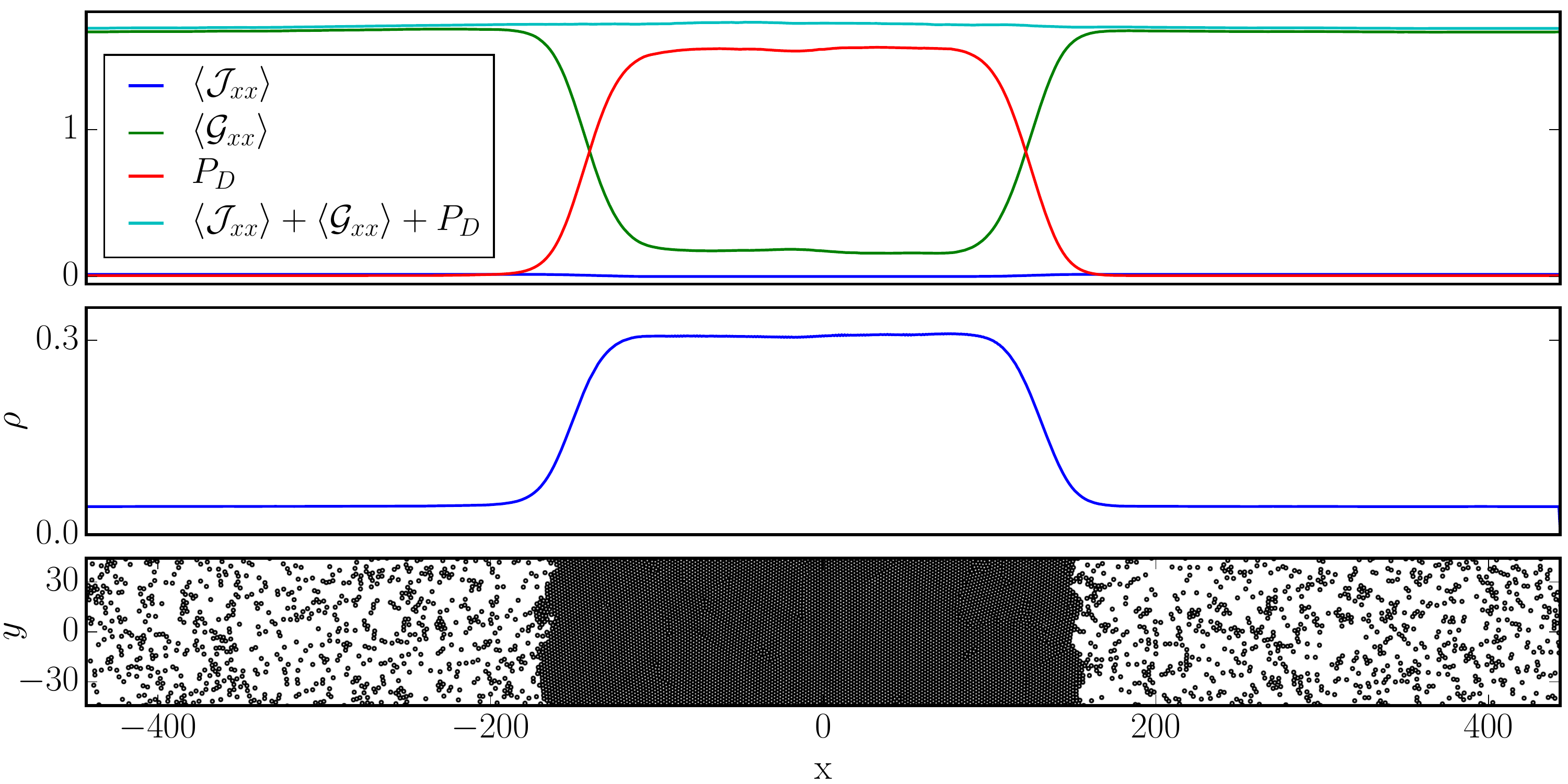}
  \end{center}
  \caption{
  Brownian dynamics simulations of $10000$ self-propelled disks with pairwise repulsive forces in a phase separated state. Periodic boundary conditions are used in both directions. The mean packing fraction is $\phi=0.4$. The repulsive potential is given by Eq.~\eqref{eq:VPFP} with $a=1$ and $k=20$. The other parameters are $f_i=1$, $D_r=0.01$, $D_t=0$, $\tilde\gamma=1$, $m=1$.
  Spatial profiles are averaged over $100$ persistence times and $100$ noise realizations starting from the equilibrated initial configuration shown in the bottom panel.
  {\bf Top}: Spatial profile of the three contributions to the pressure shown in Eq.~\eqref{eq:forcebalpair}. Their sum (light blue line) remains constant across phase boundaries. 
  {\bf Middle}: Density profile. 
}\label{fig:MIPS_PFP}
\end{figure}

To illustrate these results we have carried out numerical simulations of self-propelled particles evolving under the dynamics~\eqref{eq:manybod_eqs} with constant $f_i$. The interaction force derives from the pairwise repulsive potential
\begin{equation}\label{eq:VPFP}
V(\bfr) = \frac{k}{2} (2a-|\bfr|)^2 \ \Theta(2a-|\bfr|)
\end{equation}
corresponding to harmonic repulsion with stiffness $k$ between disks
of radii $a$. We use periodic boundary conditions in both directions,
a high Péclet number known to yield MIPS ($f_i/(\tilde\gamma a
D_r)=100$), and a slab geometry that favors vertical phase boundaries.
Figure~\ref{fig:MIPS_PFP} shows that, at steady-state, coexisting
phases have equal pressures, in accordance
with~\eqref{eq:forcebalpair}.
%

\subsection{Quorum sensing}
We now turn to the case of quorum sensing. Recall that in this
case there was no equation of state for the pressure. Using
Eq. \ref{eq:manybod_cos} and Eq. \ref{eq:xbalance}, along with the
definition given in Eq. \ref{eq:activeimpulsetensor}, we obtain for
quorum sensing particles
\begin{equation}
\label{eq:xbalance_quo}
\langle \sum_j {\bf v}_j \cdot (\nabla_{{\bf r}_j}) \frac{f_i}{D_r} \cos \theta_i \delta (x-x_i) \rangle  = \partial_x \left[ \langle {\cal J}^{x x} +{\cal G}^{ xx} \rangle \right] \;.
\end{equation} 
\begin{figure}
  \begin{center}
    \includegraphics{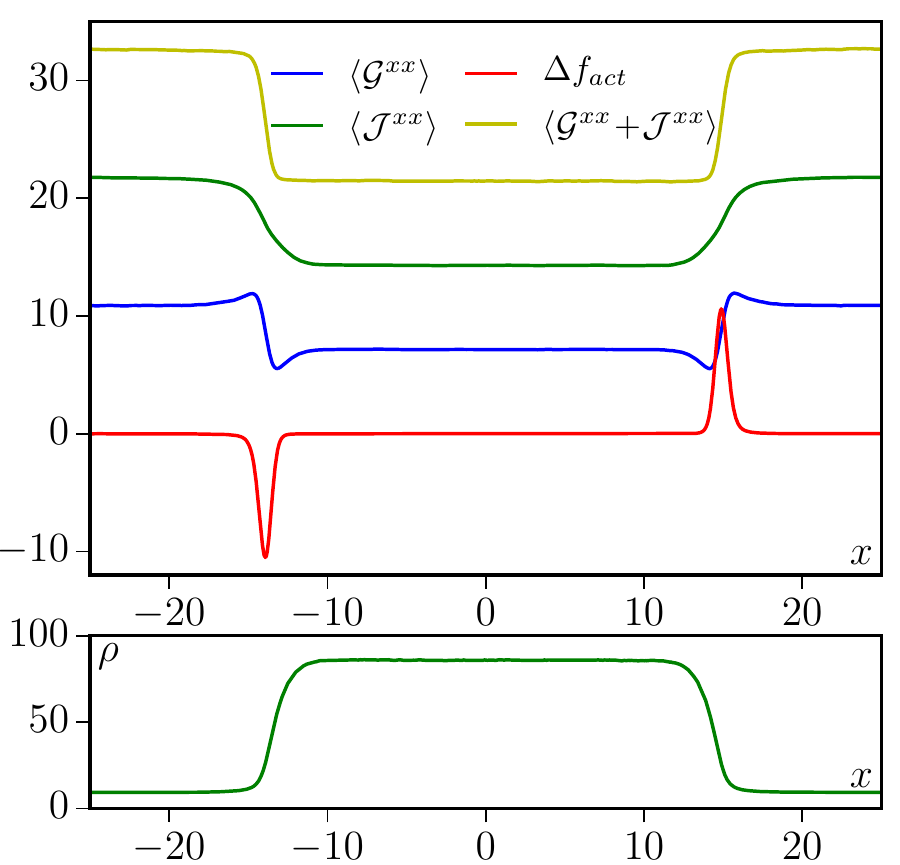}
  \end{center}
  \caption{Brownian dynamics simulations of self-propelled disks
    interacting via~\eqref{eq:QSint} and undergoing MIPS. {\bf
      Bottom:} Density profile averaged along the $\hat y$
    direction. In the steady-state, we observe the coexistence between
    a high density phase and a low-density one. {\bf Top}:
    Steady-state momentum sources and sinks $\Delta f_{\rm act}$
    localized at both interfaces make the sum of momentum and active
    impulse fluxes $\langle G^{xx}+J^{xx} \rangle$ unequal in coexisting
    phases. Parameters: $f_0=4$, $f_1=1$, $\rho_m=25$, $L_f=12.5$,
    $D_r=2$, $\gamma=1$, $dt=5.\,10^{-4}$, system size 60x10.}\label{fig:MIPSQS}
\end{figure}
At a location $x$ in the interfaces between the coexisting phases the
term $\langle \sum_j {\bf v}_j \cdot (\nabla_{{\bf r}_j})
\frac{f_i}{D_r} \cos \theta_i \delta (x-x_i) \rangle$ is in general
non-zero, much like in the vicinity of a confining walls. Moreover,
repeating the procedure leading to Eq. \ref{eq:forcebaleq} now gives
\begin{equation}
\label{eq:forcebalnce_quo}
\langle {\cal J}^{ x x}(x_\ell) +{\cal G}^{ xx}(x_\ell) \rangle +\Delta = \langle {\cal J}^{ x x}(x_r) +{\cal G}^{ xx}(x_r) \rangle
\end{equation}
with 
\begin{equation}
  \Delta=\int_{x_\ell}^{x_r} \langle \sum_j {\bf v}_j \cdot (\nabla_{{\bf r}_j}) \frac{f_i}{D_r} \cos \theta_i \delta (x-x_i) \rangle dx = \int_{x_\ell}^{x_r} \Delta f_{\rm act} dx
\end{equation}
accounting for the contribution of the steady-state momentum sources
localized at the interface. Equation~\eqref{eq:forcebalnce_quo}
relates the bulk properties of the two phases through ${\cal J}^{
  xx}$ and ${\cal G}^{ xx}$. Importantly, the relation involves
$\Delta$ which depends on the detailed structure of the interface---quite in contrast to the equilibrium case and to the case of active
particles interacting via pairwise forces. Once again, we can relate
the breakdown of standard mechanical relations to the momentum sources
measured by $\Delta f_{\rm act}$.

To illustrate these results we have carried out numerical simulations
of self-propelled disks interacting via~\eqref{eq:QSint} and
undergoing MIPS (See Fig.~\ref{fig:MIPSQS}). In the steady state, one
observes coexisting phases of low and high densities such that
Eq.~\eqref{eq:forcebalnce_quo} holds: momentum sources and sinks,
localized at the interface, make the combined flux of momentum and
active impulse $\langle {\cal J}+{\cal G} \rangle$ unequal in
coexisting phases. The lack of effective momentum conservation in the
steady-state means that coexisting phases have unequal mechanical
pressures, which contrasts with the case of pairwise forces.

\section{Conclusion}
This paper studies the existence of equations of state for the
pressure in dry active systems. Since these systems do not conserve
momentum and are out of equilibrium, the existence of an equation of
state is far from obvious.  In fact, depending on prior biases, one
might consider either the existence or absence of an equation of state
in such systems to be a surprise. Reference~\cite{Solon:2015:NatPhys}
studied several overdamped dry active systems and showed that there is
no universal answer: depending on the model, an equation of state may
or may not exist. It is therefore natural to ask for the conditions
under which an equation of state emerges.

Here, we addressed this question by considering a class of underdamped
dry active systems, from which the overdamped limit is easily
extracted. The main advantage of the underdamped model is that the
momentum field may be studied with ease allowing us to relate the
properties of the momentum field and the existence of an equation of
state. We show that, generically, the lack of momentum conservation
leads to the existence of steady-state momentum sources or sinks near
the boundaries of the system which in turn implies that an equation of
state for the pressure does not exist. Nonetheless, there is a class
of models for which an effective conservation of momentum emerges in
the steady state. For this class of systems, an equation of state
exists. The effective momentum conservation can be related to the mean
density of active forces being the divergence of the field of active
impulse, an observable that measures the momentum a particle will
receive on average from the substrate in the future. When the dynamics
of the active force leads to a vanishing mean force in the steady
state and is independent of the other degrees of freedom, the active
impulse is a local quantity, hence leading to an equation of state.

Finally, note that this paper describes only the mechanical properties
of the active particles and not the fate of the momentum source from
which they are receiving their active impulse. This is relevant for
dry systems such as bi-dimensional layers of shaken
granulars~\cite{Junot2017} or any other system in which particles are
pushing on a hard boundary~\cite{Bricard2013Nature}. When the active
particles exchange momentum with a fluid, we argue that our results
extend to the description of osmotic
pressure~\cite{Solon:2015:NatPhys,Rodenburg:2016:arxiv}, although
experiments in this case are lacking, as well as explicit treatments
from first principles of the coupling between the active particles and
the fluid.

\ack We thank M. Cates, M. Kardar, A. Morozov, J. Stenhammar for
discussions. YF was supported by the NSF, award DMR-1149266, the Brandeis
Center for Bioinspired Soft Materials, an NSF MRSEC, award DMR-1420382, and
the W. M. Keck Foundation. YF's computational resources were provided
by the NSF through XSEDE, award TG-MCB090163, and the Brandeis HPCC. 
YK is supported by an I-CORE Program of the Planning and Budgeting Committee of the Israel Science Foundation and an Israel Science Foundation grant. AS is
funded by the Betty and Gordon Moore foundation.  JT was supported by
ANR grants Bactterns.

\appendix{}

\section{Sufficient condition for the existence of an equation of state}
\label{sec:SCEOS}

In this appendix we show that if the dynamics of the active force
${\bff_i}$ of particle $i$ does not depend on the other degrees of
freedom and leads to a vanishing mean active force in the steady
state, then the pressure admits an equation of state. Note that
$\langle \bff_i\rangle=0$ in the steady state does not preclude the
existence of regions of space in which a local density of active force
$\sum_i \bff_i \delta(\bfr-\bfr_i)$ is non-zero on average. The latter
  only relies on correlations between orientations of the particles
  and their positions. 

We consider non-interacting particles evolving according to
\begin{equation}\label{eq:dyn2}
  \dot {\bf r}_i ={\bf v}_i\;;\qquad m \dot {\bf v}_i = -\tilde \gamma {\bf v}_i+{\bf f}_i -\grad_{{\bf r}_i}V_{\rm ext}+\sqrt{2 \tilde{\gamma}^2 D_t} \boldsymbol{\eta}_i.
\end{equation}
We require the evolution of the active force ${\bf f}_i$ to be
independent of the other degrees of freedom and to be given by an unspecified
master equation
\begin{equation}
  \dot P_i^f({\bf f}_i) = K_i P_i^f({\bf f}_i)
\end{equation}
for the probability $P_i^f({\bff }_i,t)$ that particle $i$ has a force
$\bff_i$ at time $t$. The fact that the dynamics of ${\bf f}_i$ is
independent of other degrees of freedom means that the operator $K_i$
solely involves ${\bf f}_i$. The joint master
equation describing the dynamics of particle $i$ is then given by
\begin{equation}\label{eq:master}
  \dot P_i(\bfr_i,\bfv_i,\bff_i,t) = - \grad_{{\bf r}_i}\!\cdot [{\bf v}_i P_i] - \grad_{{\bf v}_i}\!\cdot [-\gamma {\bf v}_i  P_i +\frac{{\bf f}_i}m P_i - \frac  {\grad_{{\bf r}_i} V_{\rm ext}}m P_i -\gamma^2 D_t \grad_{{\bf v}_i} P_i] + K_i P_i
\end{equation}

As before, we define the active impulse through
\begin{equation}\label{eq:defAIcomp}
  \bfdp(\bfr_0,\bfv_0,\bff_0,t)\equiv  \int_t^\infty \overline{\bff_i(s)} ds=  \int_t^\infty ds \int d \bff\, \bff P_i(\bff,s|\bfr_0,\bfv_0,\bff_0,t)
\end{equation}
where $P_i(\bff,s|\bfr_0,\bfv_0,\bff_0,t)$ is the probability of the
active force on particle $i$ being $\bff$ at time $s>t$, given that
the particle was at $\bfr_i=\bfr_0$ at time $t$ with velocity
$\bfv_i=\bfv_0$ and active force $\bff_i=\bff_0$. Again,
$\bfdp(\bfr_0,\bfv_0,\bff_0,t)$ is the total momentum a particle at
$\bfr_0$ with velocity $\bfv_0$ and active force $\bff_0$ will
receive, on average, during the rest of its history. Note that the
impulse is only finite if the dynamics of $\bff_i$ lead to a
vanishing mean active force in the steady state, a case to which we
restrict our discussion from now on.

We then use the fact that the dynamics of $\bff$ are homogenous in time
and do not depend on the other degrees of freedom to
rewrite~\eqref{eq:defAIcomp} in the simpler form:
\begin{equation}\label{eq:YasEq}
  \bfdp(\bff_0)\equiv  \int_0^\infty ds \int d \bff\, \bff P_i^f(\bff,s|\bff_0,0)\,.
\end{equation}
Let us now consider the action of the operator $K_i^\dagger$ on
the function $\bfdp(\bff_0)$:
\begin{eqnarray}
  K_i^\dagger \bfdp(f_0) &=& \int_0^\infty ds \int d\bff \,\bff K_i^\dagger P_i^f(\bff,s|\bff_0,0)\label{eq1}\\
  &=& \int_0^\infty ds \int d\bff\, \bff \partial_s P_i^f(\bff,s|\bff_0,0)\label{eq2}\\
  &=& - \int d\bff\, \bff  P_i^f(\bff,0|\bff_0,0)\nonumber\\
  &=& - \int d\bff\, \bff \delta(\bff-\bff_0)\nonumber\\
  K_i^\dagger \bfdp(f_0)  &=& -\bff_0\label{eq:Kd}
\end{eqnarray}
Note that $K_i^\dagger$ acts on $\bff_0$ and we used the backward
equation to go from~\eqref{eq1}to~\eqref{eq2}. Then, we consider the
evolution of the active impulse of particle $i$ over time,
$\bfdp(\bff_i(t))$. The dynamics of its average is given by
\begin{eqnarray}
\frac{d}{dt} \langle \bfdp(f_i(t)) \rangle &=&  \frac{d}{dt} \int d\bff_i \bfdp(\bff_i) P_i^f(\bff_i,t)\nonumber\\
&=& \int d\bff_i \bfdp(\bff_i) K_i P_i^f(\bff_i,t)\nonumber\\
  &=& \int d\bff_i K_i^\dagger \bfdp(\bff_i) P_i^f(\bff_i,t)\nonumber\\
  &=& \langle K_i^\dagger \bfdp(\bff_i(t)) \rangle\nonumber\\
  &=& -\langle \bff_i(t) \rangle.\nonumber
\end{eqnarray}
Similarly, we can introduce the field of active impulse density:
\begin{equation}
  \bfdpf(\bfr) = \sum_i \bfdp(\bff_i(t))\delta(\bfr -\bfr_i(t))\nonumber
\end{equation}
For non-interacting particles, the joint probability $P(\bfr_1,\bfv_1,\bff_1,\dots,\bfr_N,\bfv_N,\bff_N)$ factorizes as
\begin{equation}
  P(\bfr_1,\bfv_1,\bff_1,\dots,\bfr_N,\bfv_N,\bff_N)=\prod_{i=1}^N P_i(\bfr_i,\bfv_i,\bff_i)\nonumber
\end{equation}
so that the dynamics of $\langle \bfdpf(\bfr) \rangle$ are given by:
\begin{eqnarray}
  \partial_t\langle \bfdpf(\bfr) \rangle &=& \sum_i \int  d\bfr_id\bfv_i d\bff_i \dot P_i(\bfr_i,\bfv_i,\bff_i) \bfdp(\bff_i)\delta(\bfr-\bfr_i)\nonumber \\
  &=& \sum_i \int  d\bfr_id\bfv_i d\bff_i  \bfdp(\bff_i)\delta(\bfr-\bfr_i) \left[-\grad_{\bfr_i}\cdot \bfv_i  P_i(\bfr_i,\bfv_i,\bff_i)+K_i P_i(\bfr_i,\bfv_i,\bff_i)\right]\nonumber
\end{eqnarray}
Note that the term involving $\grad_{\bfv_i}$ in the master
equation~\eqref{eq:master} vanishes upon integration over $\bf v_i$
and that the integrals over the other particles gives $\int
d\bfr_jd\bfv_jd\bff_j P_j(\bfr_j,\bfv_j,\bff_j)=1$ . Integrating by
parts, one then gets
\begin{eqnarray}
  \partial_t \langle \bfdpf(x) \rangle &=& \sum_i \int d\bfr_i d\bfv_i d\bff_i \big[ -\grad_{\bfr} \cdot\bfv_i \bfdp(f_i)\delta(\bfr-\bfr_i) P_i(\bfr_i,\bfv_i,\bff_i) \nonumber\\&&\quad + P_i(\bfr_i,\bfv_i,\bff_i) K_i^\dagger \bfdp(\bff_i) \delta(\bfr-\bfr_i)\big]\nonumber\\
&=& -\grad_\bfr\cdot \sum_i\langle \bfv_i \bfdp(\bff_i) \delta(\bfr-\bfr_i) \rangle + \sum_i\langle K_i^\dagger \bfdp(\bff_i)\delta(\bfr-\bfr_i) \rangle\nonumber\\
&=& -\grad_\bfr \cdot\langle \sum_i\bfv_i \bfdp(\bff_i) \delta(\bfr-\bfr_i) \rangle - \langle \sum_i \bff_i\delta(\bfr-\bfr_i) \rangle\nonumber
\end{eqnarray}
where we used Eq.~\eqref{eq:Kd}. In the steady state, one thus has
\begin{equation}\label{eqapG}
  \langle \sum_i \bff_i\delta(\bfr-\bfr_i) \rangle=  -\grad_\bfr \cdot\langle \sum_i \bfv_i \bfdp(\bff_i) \delta(\bfr-\bfr_i) \rangle =  -\grad_\bfr \langle {\cal G} \rangle
\end{equation}
which then leads to an EOS as in section~\ref{sec:impulse}.

\subsection{Example with multiplicative noise}

To illustrate the above results, we consider an active force
$\bff_i=f_i \bfu_i (\theta_i)$, where
$\bfu_i(\theta_i)=(\cos\theta_i,\sin\theta_i)$ as in the main
text. Position and velocities evolve with Eq.~\eqref{eq:dyn}. The angle
$\theta_i$ undergoes rotational diffusion through the following
It\=o-Langevin dynamics:
\begin{equation}
  \dot \theta_i = \sqrt{2 D_r(\theta_i)}\, \eta_i \;.\label{eq:diffrotmu}
\end{equation}
which leads to the Fokker-Planck equation
\begin{equation}
\partial_t P_i^f(\theta_i) = \frac{\partial^2}{\partial \theta_i^2} D_r(\theta_i)
P_i^f(\theta_i)\nonumber
\end{equation}
Here, $K_i^\dagger=D_r(\theta_i)\displaystyle \frac{\partial^2}{\partial \theta_i^2}$ so that the impulse can be computed from Eq.~\eqref{eq:Kd} through
\begin{equation}
\frac{\partial^2}{\partial \theta_i^2}  {\bf \bfdp}(\theta_i) = -\frac{f_i}{D_r(\theta_i)} {\bf u}(\theta_i)\
\end{equation}
which we solve for a given example below.
\begin{figure}
  \begin{center}
  \includegraphics[width=.4\textwidth]{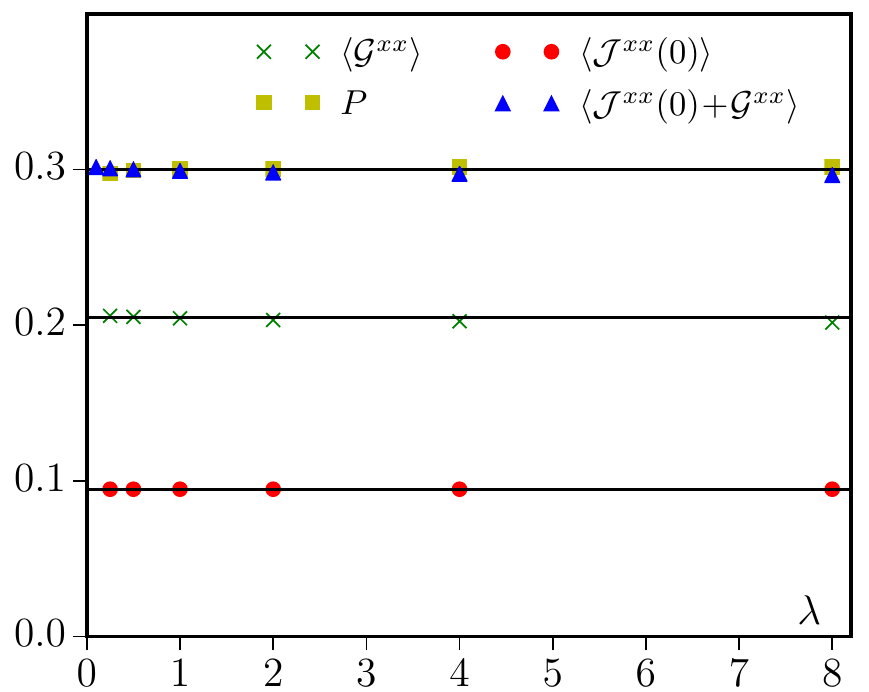}
  \includegraphics[width=.4\textwidth]{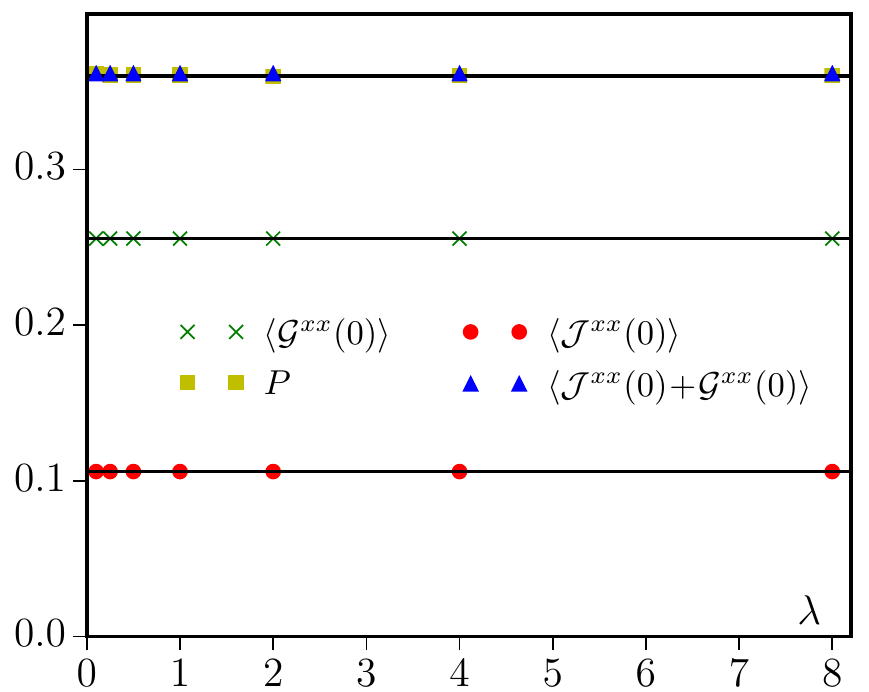}
\end{center}
\caption{Brownian dynamics simulations of self-propelled disks
  undergoing dynamics~\eqref{eq:dyn} and~\eqref{eq:diffrotmu} for
  $m=1$, $\tilde\gamma=2$, $f_i=1$, $\rho_0=1$,
  $dt=5.\,10^{-3}$, $L_x=20$.  Straight lines are guides for the eyes
  and show that this system also admits an equation of state for the
  mechanical pressure. To measure ${\cal G}^{xx}(0)$, we use
  Eq.~\eqref{eqapG}.  {\bf Left:} Here, the active impulses of the
  particles, $\bfdp$, are measured using the
  definition~\eqref{eq:defAIcomp} and $\epsilon=0.2$. {\bf Right:}
  Here, the active impulse of the particles, $\bfdp$, are measured
  from~\eqref{eq:AI} and $\epsilon=0.4$.  }\label{fig:EOSMN}
\end{figure}

Following the previous section or directly using It\=o calculus, it is
easy to show that the dynamics of the density of free active impulse
is then given by
\begin{equation}
\partial_t \langle \sum_i {\bf \bfdp}(\theta_i(t)) \delta({\bf r}-{\bf r}_i)
\rangle = - \langle \sum_i f_i {\bf u}(\theta_i(t)) \delta({\bf r}-{\bf r}_i)
\rangle - \nabla \cdot \langle\sum_i {\bf v}_i {\bf \bfdp}(\theta_i(t))\delta({\bf
r}-{\bf r}_i)\rangle
\end{equation}
so that, in steady-state, the local force density is given by the
divergence of the flux of active impulse
\begin{equation}
\langle \sum_i f_i {\bf u}(\theta_i(t)) \delta({\bf r}-{\bf r}_i) \rangle =  -
\nabla \cdot \langle {\cal G} \rangle; \qquad {\cal G} = \sum_i {\bf v}_i {\bf
\bfdp}(\theta_i(t))\delta({\bf r}-{\bf r}_i) \,.
\end{equation}

For concreteness, we consider 
\begin{equation}
  D_r(\theta) = \frac{1}{1+\eps \cos(2\theta)}
\end{equation}
which is such that the steady-state $P(\theta)$ is anisotropic but
such that $\langle {\bf u}(\theta)\rangle=0$. Then,
the equation for $\bfdp(\theta)$ is
\begin{equation}
\partial_{\theta_i}^2  {\bf \bfdp}(\theta_i) = f_i (1+\eps \cos(2\theta_i)) {\bf u}(\theta_i)
\end{equation}
whose solution is
\begin{equation}\label{eq:AI}
  {\bf \bfdp} = f_i {\bf u}(\theta_i) + f_i \frac{\eps}2 \left(\begin{array}{l cr} \cos \theta_i+\frac 1 9 \cos3\theta_i \\ -\sin\theta_i+\frac 1 9 \sin3\theta_i
  \end{array}\right)
\end{equation}
Fig~\ref{fig:EOSMN} shows that, indeed, the pressure is given in this
case by $P=\langle G^{xx}({x_b})+J^{xx}({x_b}) \rangle$ with ${x_b}=0$, hence satisfying an equation of state: it is independent of the stiffness of the confining potential.

\vspace{.25cm}
\begin{center}
\rule{0.75\textwidth}{.4pt}
\end{center}
\vspace{.25cm}

\bibliography{biblio}

\end{document}